# Integrating transfer matrix method into SCAPS-1D for addressing optical losses and per-layer optical properties in perovskite/Silicon tandem solar cells

Peymaneh Rafieipour*, Aminreza Mohandes, Mohammad Moaddeli, Mansour Kanani

*Department of Materials Science and Engineering, School of Engineering, Shiraz University, Shiraz, Iran*

*Corresponding author: prafieepoor@yahoo.com*

**Abstract**

SCAPS-1D software ignores optical losses and recombination junction (RJ) layer in studying tandem solar cells (TSCs). This paper presents an optoelectronic study of a perovskite/Silicon TSC, comparing the effects of using two different methods of calculating filtered spectra on the photovoltaic performance parameters of tandem device. It is shown that integrating transfer matrix (TM) method into SCAPS-1D addresses per-layer optical losses and provides a platform for optimizing the RJ layer in TSCs. Using Beer-Lambert (BL) method for calculating the filtered spectra transmitted from the perovskite top sub-cell is revealed to overestimate the cell efficiency by ~4%, due to its inability to fully address optical losses. Also, the BL method fails to tackle any issues regarding optical improvement through ITO ad-layer on the RJ. Using TM formalism, the efficiency of the proposed perovskite/Silicon TSC is shown to be increased from 19.81% to 23.10%, by introducing the ITO ad-layer on the RJ. It is the first time that the effect of filtered spectrum calculation method is clearly investigated in simulating TSCs with SCAPS-1D. The results pave the way to introduce the optical loss effects in SCAPS-1D and demonstrate that the BL method that has been used before needs to be revised.

## 1- Introduction

Since their first fabrication in 1955 [1], silicon solar cells (SSCs) have demonstrated power conversion efficiencies (PCEs) over 25% [2]. In spite of their high stability and advanced processing technology, their efficiencies are still facing challenges in terms of Shockley-Queisser (SQ) theoretical limit [3]. The easiest way to overcome the SQ limitation of a solar cell (SC) device involves designing a series of devices called tandem solar cells (TSCs), by connecting two or multiple SCs that are electrically or mechanically linked together [4]. 2-terminal (2-T) monolithic TSCs are the most attractive and cost-effective architectures in which the layers are deposited directly on top of each other and the device requires only one external circuit and one substrate. To attain high efficiencies in 2-T monolithic TSCs, the currents between two sub-cells are required to become matched and the optical losses are required to become minimized [4]. The electrical coupling of top and bottom sub-cells linked in series in 2-T monolithic TSCs is provided by the recombination junction (RJ) that could be comprised of transparent conducting oxides (TCOs) such as ITO as the recombination layer or two adjacent heavily n-doped ($n^{++}$) and p-doped ($p^{++}$) regions [5]. High optical transmittance and electrical conductivity as well as low lateral electrical conductivity make TCOs as ideal candidates for the recombination layer in 2-T TSCs [5]. However, they pose reflection losses due to their refractive index mismatch with adjacent layers. Therefore, light management is a coupled requirement in order to form an appropriate 2-T monolithic TSC [6]. On the other hand, selecting suitable top and bottom sub-cells are another major factor that should be considered for breaking the SQ efficiency limit.

Organic-inorganic lead halide perovskites (PVKs) are excellent absorber materials for SCs due to their extraordinary properties such as high optical absorption coefficient, long carrier diffusion length, long carrier lifetime and tunable band gap [7]. The PCE of perovskite SCs (PSCs) has been increased from 3.8% in 2009 to 25.7% in 2022, approaching the record of the leading silicon reputable technology [8]. The high band gap of PVKs makes them the best candidate for the top sub-cell to be linked with silicon bottom sub-cell in 2-T monolithic TSCs, as demonstrated for the first time in 2015 [9]. To achieve the highest performance for 2-T monolithic Perovskite/Silicon TSCs, modifications to the configuration as well as materials have been reported [8]. Notably, the highest efficiency of a 2-T monolithic Perovskite/Silicon TSC was accomplished by KAUST (King Abdullah University of Science and Technology) and was reported as 33.7% [https://www.nrel.gov/pv/cell-efficiency.html].

With the continuity and growing research interest in TSCs, more accurate and realistic performance prediction modeling of them is in demand. Solar Cell Capacitance Simulator (SCAPS) software is a well-known powerful software introduced for simulating different SC devices and plotting the results in the form of energy band diagrams, current density-voltage (J-V) curve, external quantum efficiency (EQE) curve and performance parameters [10]. To obtain J-V curve of the SC, SCAPS-1D solves Poisson and continuity equations with exact boundary conditions for holes and electrons in one dimension (see tables S1-S5). However, SCAPS-1D faces major challenges for simulating 2-T monolithic TSCs due to the existence of RJ layer. A common approach for solving this issue is to study the top sub-cell under the illumination of AM1.5 Global (AM1.5G) sun spectrum and the bottom sub-cell under the illumination of the filtered spectrum transmitted from the top sub-cell. In many previous theoretical studies of 2-T monolithic TSCs using SCAPS-1D, the Beer-Lambert (BL) method was used for calculating the filtered spectra [11-23]. In 2021, P. Yang et. al. evaluated the potential of using $CsPb(I_{1-x}Br_x)_3$ perovskites as the wide bandgap top sub-cell along with the low bandgap crystalline silicon bottom sub-cell in two-terminal (2-T) and four-terminal (4-T) tandem solar cells [11]. They obtained the optimum perovskite bandgap required to maximize efficiency in each configuration. In 2-T and 4-T tandem configurations, it was demonstrated that the maximum efficiencies of 29.23% and 28.5% can be achieved by $CsPbI_3$ and $CsPbBr_3$ top sub-cells, respectively. Also, the optimized perovskite thicknesses were obtained as 275 nm and 400 nm in 2-T and 4-T tandem configurations, respectively [11]. In another theoretical study, J. Madan et. al designed a 23.36% perovskite-PbS CQD tandem device with 1.79 V (Voc), 16.67 mA.cm$^{-2}$ (Jsc) and 78.3% (FF) performance photovoltaic parameters [12]. They obtained the conversion efficiencies of 14.60% and 9.07% for top and filtered bottom sub-cells with optimized thicknesses of 143 nm and 470 nm, respectively. They changed the $MAPbI_3$ perovskite absorption coefficient, perovskite thickness and perovskite/transport layer defect densities, evaluating their impacts on the perovskite top sub-cell performance parameters and transmitted filtered spectra [12]. In 2021, N. Shrivastav et. al. optimized a 29.15% efficient perovskite-silicon TSC to unlock 33% conversion efficiencies [13]. They varied the thicknesses of top and bottom sub-cells simultaneously to obtain the optimized thickness values required for current matching. The perovskite top and silicon bottom sub-cells were simulated under the illuminations of AM1.5G and filtered spectra, respectively. They achieved the conversion efficiencies of 20.58% and 12.15% for top and bottom sub-cells with optimized thicknesses of 336 nm and 150 micron, respectively. The photovoltaic parameters of the optimized tandem device were reported as 2.02 V (Voc), 20.11 mA.cm$^{-2}$ (Jsc), 81.36% (FF) and 33.05% efficiency [13]. Very recently, N. Shrivastav et. al. studied a perovskite/CIGS monolithic TSC and optimized the thicknesses of top and bottom sub-cells to achieve the 1.92 V (Voc), 20.04 mA.cm$^{-2}$ (Jsc), 77% (FF) and 29.7% efficiency for a tandem device [14]. The matched current density was reported for the absorber thicknesses of 347 nm and 2 micron for top and bottom sub-cells, respectively [14]. In 2022, Jafarzadeh et. al. designed a perovskite-homojunction SnS tandem solar cell and achieved the optimized 28.92% efficiency [15]. They started with the optimization of SnS homojunction SC by varying layer thicknesses, doping concentrations, defect densities and interface defects. Then, an optimization of the perovskite SC by varying the absorber bandgap and thickness was performed. Thereafter, the tandem device was simulated with the optimized bandgap of 1.67 eV for the perovskite layer and the photovoltaic parameters of 1.99 V (Voc), 16.99 mA.cm$^{-2}$ (Jsc), 85.15% (FF) and 28.92% efficiency were obtained following the current-matching technique [15].

In the aforementioned articles and other related researches published in the literature [11-23], increasing the efficiency of TSC was achieved through choosing the suitable active material/transport layers and optimizing material characteristics. Few researches have paid attention to this important point that the BL method used for calculating the filtered spectra does not include all optical loss mechanisms. There are different optical losses affecting the transmitted light spectrum, including parasitic light absorption in charge transport layers, light reflections from the interfaces between adjacent layers and light scattering from the grain boundaries inside the layers. Since the performance of the bottom sub-cell is highly dependent on the filtered spectrum, it is necessary to give more importance to the method of calculating the filtered spectrum. Although the BL method is very simple and straightforward, it is only valid for homogeneous media, under some approximations that will be addressed in the following. First, light scattering in the layers is ignored since the real part of the complex refractive indices of materials is assumed to be zero. Second, the light reflection from the interfaces between different layers is ignored because the refractive index mismatch between the adjacent layers and the Fresnel reflection coefficient equals zero. Therefore, the reported values for the photovoltaic parameters and optimized thicknesses of the tandem SCs are overestimated and not reliable for practical applications. This necessity demands that a research work with an optical view deals with the simulation of tandem SCs using SCAPS-1D. Up to our knowledge, there is only one paper that has used TM method for calculating the filtered spectra and simulating a perovskite-silicon tandem solar cell with SCAPS-1D. In 2023, R. Pandey et. al. designed a perovskite-Silicon tandem solar cell with the use of a lead-free MASnI$_2$Br$_1$ perovskite and achieved a maximum conversion efficiency of 30.7% with a V$_{oc}$ of 2.14 V [24]. They reported a performance investigation of MASnI$_{3-x}$Br$_x$-based perovskite SCs by varying halide composition, perovskite thickness and perovskite bulk defect density [24]. Although the results were interesting, the effects of using different methods for calculating the filtered spectrum on the photovoltaic performance parameters of the tandem SC were not discussed. Therefore, the importance of using the right method in the simulation of 2-terminal monolithic TSCs with SCAPS-1D is still not clear. To fulfill this requirement, in this paper, we aim to evaluate the performance parameters of an emerging perovskite-silicon tandem SC using TM and BL methods and compare their corresponding results with each other. Standalone and 2-T monolithic tandem configuration of the Perovskite/Silicon as top/bottom sub-cells will be accomplished by absorber layer thickness variation, energy band diagram, J–V curve, EQE curve, filtered spectra, current matching, and tandem performance parameters, using SCAPS-1D. In addition, optical loss effects related to two commonly used RJs including: (I) Spiro-OMeTAD (HTL)/$n^{++}$-Silicon hybrid RJ and (II) Spiro-OMeTAD (HTL)/ITO/$n^{++}$-Silicon RJ will be studied with SCAPS-1D. Using TM method, it will be shown that the device performance metrics is improved by inserting an ITO layer in the RJ, as was expected from the literature [25]. In addition, we will analyze per-layer parasitic light absorption and total reflection from perovskite top sub-cell, by using TM method. We will show that the major contribution of total optical loss is from interface reflections rather than the layer parasitic absorption or intralayer light scattering. That is why BL method fails to fully address optical losses and results in an overestimation of the cell efficiency.

The remaining parts of the paper are organized as follows: The next section covers the device structures, simulation methodology and methodology dependence of filtered spectra for both TSC configurations with different RJs. The formulations of TM and BL methods will be described in sub-section 2-2. In the third section, simulation results with using both TM and BL methods will be presented and the effects of including optical loss effects in SCAPS-1D will be discussed.

## 2- Device structure and simulation methodology

### 2-1 SCAPS-1D device simulation

The present work is carried out using SCAPS-1D software, developed in the Department of Electronics and Information Systems (ELIS), University of Gent, Belgium [26]. The proposed 2-T monolithic tandem architecture is composed of perovskite and silicon SCs that are stacked on each other. The simulation methodology of TSCs through SCAPS-1D concerns two steps: First, standalone simulations of perovskite top sub-cell and silicon bottom sub-cell are carried out under the illumination of AM1.5G spectrum to optimize the PV parameters of each device. Next, the PVS/Silicon TSC is simulated by the current matching strategy, using the AM1.5G and filtered spectra illuminated on the perovskite top sub-cell and silicon bottom sub-cell, respectively. Figure 1(a) and (b) present a schematic illustration of the standalone simulations of perovskite top sub-cell and silicon bottom sub-cell, respectively. The top sub-cell is composed of fluorine-doped tin oxide (FTO) as the front contact, $SnO_2$ as the electron transport layer (ETL), perovskite as the absorbing layer, Spiro-OMeTAD as the hole transport layer (HTL) and Au as the back contact. In addition, the bottom sub-cell consists of $n^{++}$-Si as the ETL, n-Si as the absorbing layer and $p^{++}$-Si as the HTL. The layer materials and their corresponding thicknesses are described in Figure 1(a) and (b). When studying perovskite/Silicon TSCs, which are illustrated schematically in Figure 1(c) and (d), the filtered spectrum transmitted from the perovskite top sub-cell is illuminated on two different configurations of the bottom sub-cell that are different in their RJ. In addition, the thickness of HTL ($n^{++}$-Si) in each configuration has been optimized, independently. The perovskite active layer is $Cs_{0.05}(FA_{0.85}MA_{0.15})_{0.95}Pb(I_{0.85}Br_{0.15})_3$, which is chosen due to its large band gap (1.57-1.60 eV) [27] and denoted by CsFAMA in this paper. In the simulation, CsFAMA is assumed as a single-graded perovskite with a linearly graded band gap varied between 1.57 eV and 1.60 eV.

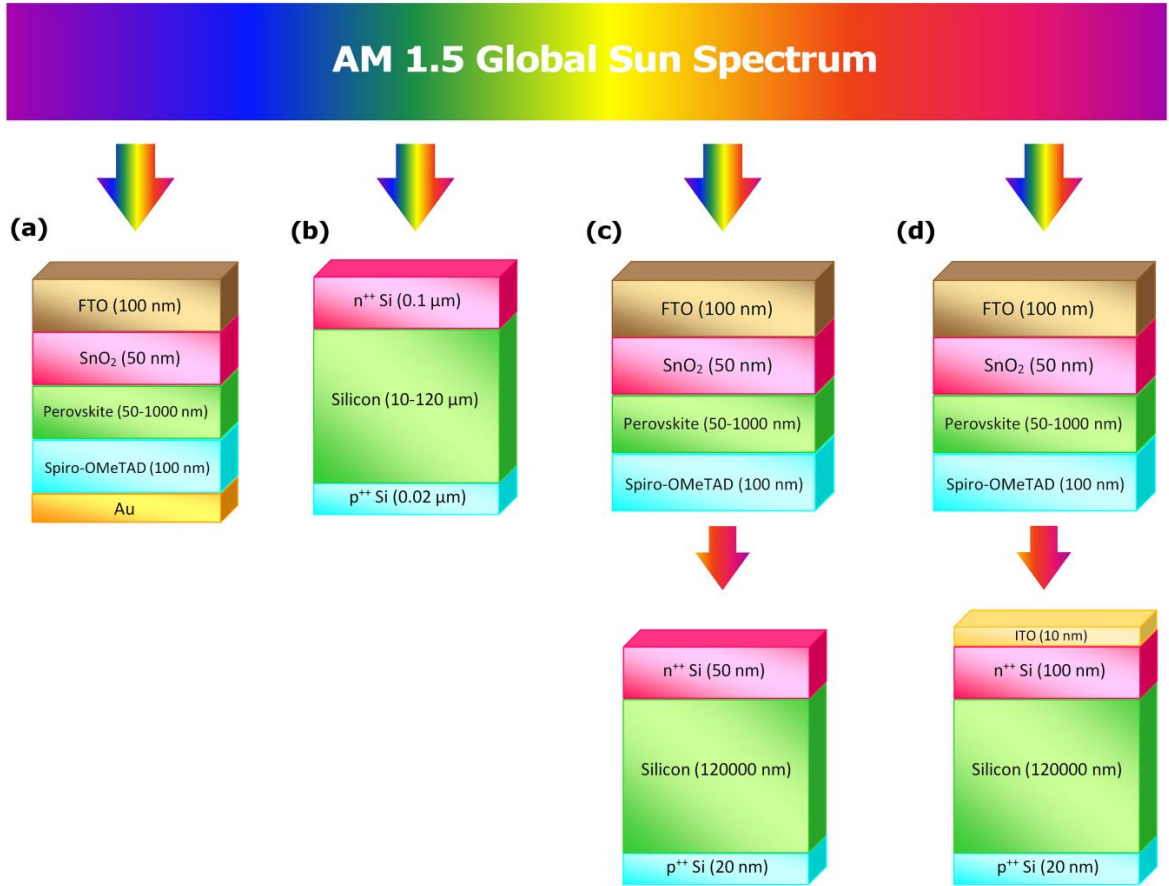

Figure 1: Schematic illustration of the SCAPS-1D simulation methodology: (a) standalone PSC under AM1.5G light irradiation, (b) standalone SSC under AM1.5G light irradiation, and (c) 2-T monolithic Perovskite/Si TSC with AM1.5G spectrum illuminates the top sub-cell and the filtered spectrum illuminates the bottom sub-cell which is designed in two different configurations.

To obtain the required parameters for simulating PSC, they are calibrated in such a way to reproduce the experimental J-V curve obtained by Bu et. al. [28]. According to [28], the thicknesses of $SnO_2$, CsFAMA and Spiro-OMeTAD layers are assumed as 50 nm, 700 nm and 300 nm, respectively. Figure 2 shows that the calibrated results are in accordance with the experimental J-V curve of the PSC [28], which confirms the validity of our calibration of the top sub-cell design. We use the SSC, studied in [16], as the bottom sub-cell in the designed perovskite/silicon tandem configuration.

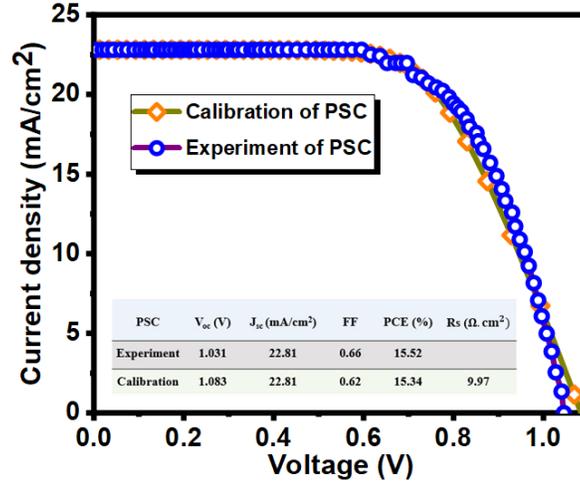

Figure 2: The comparison between the experimental J-V curve reported in [28] and the obtained calibrated data for the J-V curve of the PSC with the CsFAMA absorbing layer. $R_s$ is the series resistance.

The next step is to optimize the top and bottom sub-cells in standalone conditions. To minimizing parasitic absorption losses (see Figure S1); the thicknesses of $SnO_2$ and Spiro-OMeTAD are chosen as 50 nm and 100 nm, respectively [29,30]. The thickness of CsFAMA is then varied and J–V curve as well as EQE curve of the PSC is simulated. The results are shown in Figure 3. An optimization study of PV parameters of the PSC demonstrates the PCE of 15.68%, $V_{oc}$ of 1.10 V, $J_{sc}$ of 21.97 mA/cm$^2$, FF of 64.97% and the series resistance ($R_s$) of 9.97 Ω.cm$^2$. Similar procedure is done for the silicon bottom sub-cell and the results are depicted in Figure S8. Also, the dependency of the performance parameters $V_{oc}$, $J_{sc}$, FF, and PCE of the top and bottom sub-cells on the thicknesses of their corresponding absorber layers are illustrated in Figures S6 and S9.

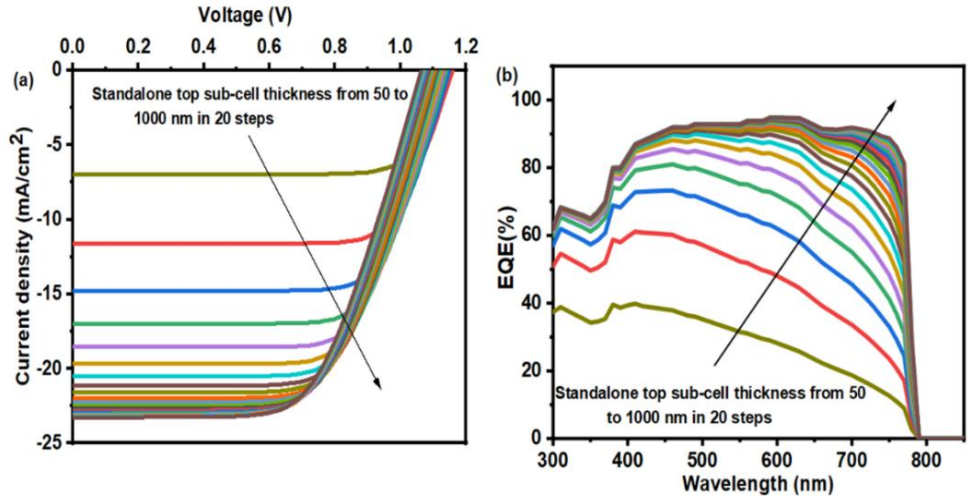

Figure 3: (a) J-V, and (b) EQE curves of the perovskite top sub-cell under the thickness variation of CsFAMA absorbing layer from 50 to 1000 nm, in 20 steps. Standalone calibration is done under the illumination of AM1.5G spectrum.

## 2-2 Formulations of TM and BL methods

To calculate the filtered spectrum, the perovskite top sub-cell is assumed as a multilayer structure composed of m individual layers and the transmitted light is obtained by TM and BL methods. In the BL method, the filtered spectrum is computed by the following equation [11-23]:

$$I(\lambda) = I_0(\lambda) \exp\left(\sum_{j=1}^{m} -\alpha_j(\lambda) d_j\right) \qquad (1)$$

where $I_0(\lambda)$ denotes the AM1.5G spectrum, $\lambda$ is the wavelength, $\alpha_j(\lambda)=4\pi\kappa_j/\lambda$ is the absorption coefficient of each layer in the perovskite top sub-cell, $\kappa_j$ is the extinction coefficient and $d_j$ is the layer thickness. On the contrary, in the TM method, a total TM (S) is assigned to the multiplayer structure that connects the amplitudes of the electric fields on both sides of it [31]:

$$\begin{bmatrix} E_0^+ \\ E_0^- \end{bmatrix} = S \begin{bmatrix} E_{m+1}^+ \\ E_{m+1}^- \end{bmatrix} \qquad (2)$$

where $E^+$ and $E^-$ refer to the amplitudes of the electric fields propagating in the direction of light incidence and against it, respectively. The index 0 and m+1 are used to denote the incident and outgoing media, respectively. For a stack of linear, homogeneous and isotropic media, the total TM is a 2×2 matrix calculated by multiplying the transfer matrices of each layer. The individual transfer matrices of the layers are obtained by multiplying two other matrices named as phase matrix I and refraction matrix L that describe optical characteristics of the layers.

$$S = \begin{bmatrix} S_{11} & S_{12} \\ S_{21} & S_{22} \end{bmatrix} = \left(\prod_{j=1}^{m} I_{(j-1)j} L_j\right) \cdot I_{m(m+1)} \qquad (3)$$

When light travels through a multilayer structure, it encounters successive refraction at the interfaces and propagation inside the layers. The refraction matrix accounts for the light reflection and transmission at the interfaces between two adjacent layers. Considering the light refraction at the interface between the jth layer and the (j+1)th, the refraction matrix is defined as [31]:

$$I_{j(j+1)} = \frac{1}{t_{j(j+1)}} \begin{bmatrix} 1 & r_{j(j+1)} \\ r_{j(j+1)} & 1 \end{bmatrix} \qquad (4)$$

where $r_{jk}$ is the complex Fresnel reflection coefficient and $t_{jk}$ is the complex Fresnel transmission coefficient at the interface between the jth layer and the (j+1)th layer. At normal incidence, the Fresnel reflection and transmission coefficients are [31]:

$$r_{j(j+1)} = \frac{N_j - N_{(j+1)}}{N_j + N_{(j+1)}} \qquad (5)$$

$$t_{j(j+1)} = \frac{2N_j}{N_j + N_{(j+1)}} \qquad (6)$$

The phase matrix describes the propagation of the electric field inside the medium and is defined as [31]:

$$L_j = \begin{bmatrix} e^{-i\delta_j d_j} & 0 \\ 0 & e^{i\delta_j d_j} \end{bmatrix} \quad (7)$$

where $d_j$ is the thickness of the jth layer, and $\delta_j = 2\pi N_j/\lambda$ is the phase change of the electromagnetic wave when it travels inside the jth layer. $N_j$ is the complex refractive index of the jth layer and is given by $N_j = n_j + i\kappa_j$, where $n_j$ and $\kappa_j$ are the index of refraction and the extinction coefficient of the jth layer. The total complex reflection and transmission coefficients of the multilayer structure can then be calculated from the total S matrix by the following equations [31]:

$$r = \frac{E_0^-}{E_0^+} = \frac{S_{21}}{S_{11}} \quad (8)$$

$$t = \frac{E_{m+1}^+}{E_0^+} = \frac{1}{S_{11}} \quad (9)$$

The transmitted and reflected intensities from the multilayer structure are then calculated by two physical quantities, total transmittance (T) and total reflectance (R), defined as follows [31]:

$$T = \frac{I_{tra}}{I_0} = \frac{n_{m+1}}{n_0}|t|^2 \quad (10)$$

$$R = \frac{I_{ref}}{I_0} = |r|^2 \quad (11)$$

where $n_0$ and $n_{m+1}$ are the refractive indices of the incident and outgoing media, respectively. Due to the energy conservation, the total absorbance is defined as:

$$A = 1 - T - R \quad (12)$$

As derived from the aforementioned equations, understanding the complex refractive index of the layers is necessary for obtaining the optical properties of the multilayer structure. In this paper, the complex refractive indices of FTO, $SnO_2$, CsFAMA, Spiro-OMeTAD, crystalline silicon and ITO are derived from the literature [32-37]. Also, the AM1.5G sun spectrum (1000 W/m$^2$), defined in the SCAPS-1D library, is used as the light illuminated on the perovskite top sub-cell.

## 2-3 Top cell filtered spectrum dependency on BL and TM method

Figure 4 shows the filtered spectra for two different configurations of silicon bottom sub-cell without and with the ITO recombination layer, respectively. The perovskite thickness is 100 nm. The transmitted spectra in each case are calculated by using BL and TM methods, via the equations 1 and 11, respectively. Using TM method, it is observed from Figure 4(a) that less light is transmitted from top perovskite sub-cell in the visible and near-infrared (NIR) wavelengths. The reason is referred to the interfacial reflection and scattering losses that are incorporated in the formalism of TM method, compared with BL method. It is then inferred that the PV parameters of the bottom sub-cell as well as the Perovskite/Silicon TSC are dependent on the method of calculating the filtered spectrum. This will be discussed later in sub-section 3-1. In addition, the AM1.5G sun spectrum and the perovskite thickness variation of the filtered spectrum are shown in Figures S10 and S11, using TM and BL methods, respectively. It is inferred from these figures that increasing the perovskite thickness enhances the light absorption in the visible part of the spectrum and results in the reduction of the transmitted light.

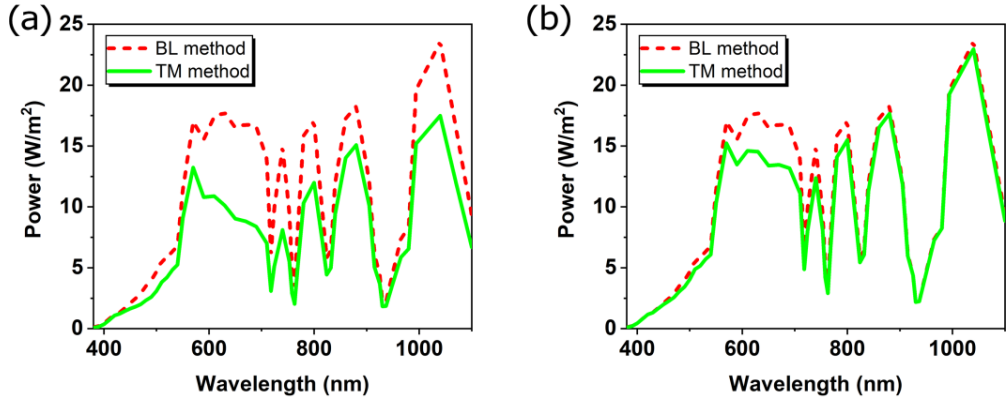

Figure 4: The filtered spectra corresponding to the perovskite thickness of 100 nm for two TSC configurations with RJs of: (a) Spiro-OMeTAD/ $n^{++}$-Si RJ, and (b) Spiro-OMeTAD/ITO/$n^{++}$-Si.

Using TM method, it is also demonstrated from Figure 4(b) that introducing ITO as the recombination layer to the RJ enhances the light transmitted from the perovskite top sub-cell. On the contrary, using BL method for calculating the filtered spectrum, it is shown by comparing the red dashed lines presented in Figure 4(a) and (b) that the filtered spectrum exhibits no difference with the addition of ITO recombination layer. Therefore, BL method is unable to reveal the effects of modifying and optimizing the RJ in SCAPS-1D. It is because any information of the outgoing medium including its refractive index is not incorporated in the BL formalism, as shown in equation 1. However, by looking at equations 5, 6 and 10 in the TM formalism, it is revealed that the refractive index of the outgoing medium (with the index of m+1) affects the amount of light transmitted and reflected. As it is stated by equation 5, the lower the refractive index contrast between two adjacent layers, the lower the light reflection coefficient from the interface between the layers. Figure 5(a) and (b) show the refractive index of Spiro-OMeTAD [36] in the comparison with that of ITO [35] and crystalline silicon [32], respectively. Since the difference between the refractive index values of ITO and Spiro-OMeTAD is lower at wavelengths above 700 nm (NIR wavelengths) compared with that of $n^{++}$-Si and Spiro-OMeTAD, the fraction of light reflected from the Spiro-OMeTAD/ITO interface is lower than that from the Spiro-OMeTAD/$n^{++}$-Si interface. Hence, the optical loss in the case where the ITO layer is added to the RJ is decreased and the light transmitted from the perovskite top sub-cell is increased at NIR wavelengths. Due to the increased optical power illuminated on the silicon bottom sub-cell, its PV parameters are expected to be enhanced in the case where the ITO layer is sandwiched between Spiro-OMeTAD and $n^{++}$-Si layers. The effects of introducing the ITO recombination layer to the RJ on the PV parameters of the silicon bottom sub-cell and the perovskite/Slicon TSC will be discussed in sub-section 3-2.

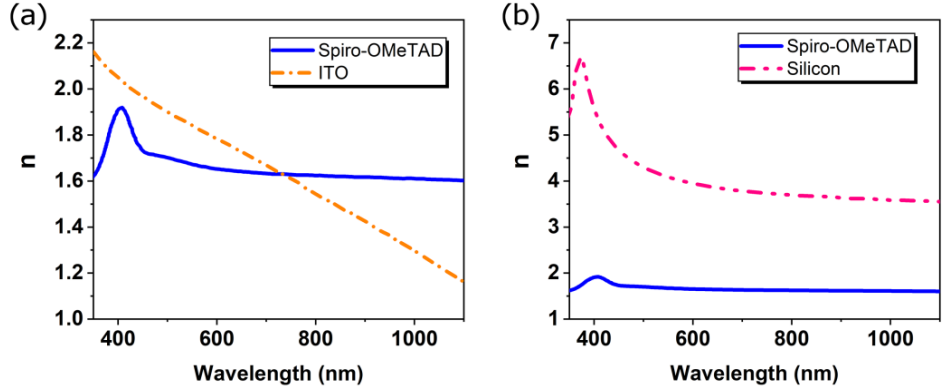

Figure 5: Comparing the refractive indices of: (a) Spiro-OMeTAD with ITO, (b) Spiro-OMeTAD with silicon ($n^{++}$-Si). Data are extracted from the literature [32,35,36].

## 3- Result and discussion

### 3-1 Photovoltaic parameters sensitivity on BL vs. TM

In order to evaluate the methodology dependence of the optimized PV results of the 2-T monolithic perovskite/Silicon TSC, first, it is needed to find the optimized thicknesses at which a same current pass through the top and bottom sub-cells. Therefore, we change the perovskite thickness and calculate the corresponding filtered spectra transmitted from the top perovskite sub-cell. Each filtered spectra is then illuminated on the silicon bottom sub-cell and its resultant Jsc values for different silicon thicknesses are obtained by SCAPS. Figure 6 depicts the $J_{sc}$ variation of the bottom sub-cell as a function of the silicon thickness, when TM method is used for calculating different filtered spectra corresponding to different perovskite thicknesses. Here, the RJ is formed by Spiro-OMeTAD (with 100 nm thickness and $N_A$ of $10^{18}$ $cm^{-3}$) HTL of top sub-cell and $n^{++}$-Si (with 100 nm thickness and $N_D$ of $10^{22}$ $cm^{-3}$) ETL of bottom sub-cell, when stacked on each other. It is observed that regardless of the perovskite thickness, $J_{sc}$ of the bottom sub-cell is increased to a maximum value and then decreased. It is attributed to the ability of charge carries to travel across the silicon absorber layer and reach the back electrode [38]. When the silicon thickness is very high, it is probable that many charge carriers recombine before entering the charge extraction layer. Hence, at higher values of the silicon thickness, less charge carriers can reach the back electrode and $J_{sc}$ decreases. Another noticeable result is that the maximum value of $J_{sc}$ passing through the bottom sub-cell is decreased by increasing the perovskite thickness. It is due to the enhanced absorption of light in the perovskite layer that decreases the light intensity illuminated on the bottom sub-cell (see Figures S10 and S11). Furthermore, the silicon thickness at which the maximum $J_{sc}$ is achieved shifts to higher values by an increase in the perovskite thickness. In other words, from a practical point of view, a higher silicon thickness is needed for a higher perovskite thickness in order to attain a high value of $J_{sc}$. It is then inferred from the obtained results that the maximum value of 120 micron for the silicon thickness is sufficient for optimizing the PV parameters of the 2-T monolithic perovskite/Si TSC in the current matching technique. Therefore, in the following, we will increase the silicon thickness up to 120 microns to have an increasing increment for the $J_{sc}$ of the bottom sub-cell and avoid its reduction. The detailed Jsc values of the bottom sub-cell corresponding to different silicon thicknesses are presented in table S6. A same behavior (not shown here) is obtained when the BL method is used for calculating the filtered spectra.

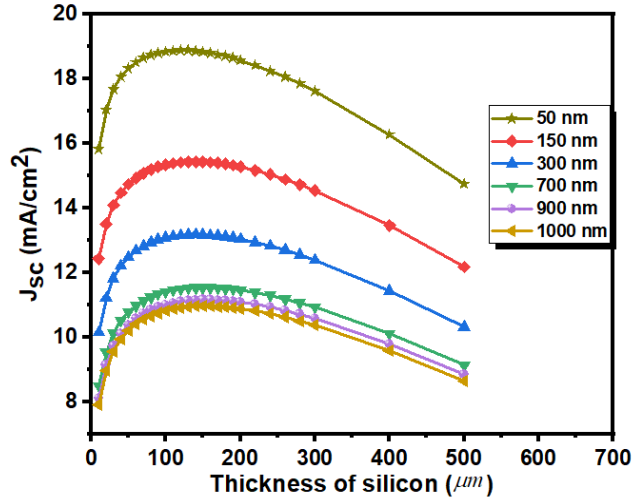

Figure 6: $J_{sc}$ of the bottom sub-cell against the thickness of the silicon layer for different perovskite thicknesses of 50, 150, 300, 700, 900, 1000 nm. The TM method is used for calculating the filtered spectra.

Figure 7 (a) to (d) provide 2D counter plots of $J_{sc}$, PCE, FF, and $V_{oc}$ of the bottom sub-cell, describing their variations as a function of perovskite and silicon thicknesses. The results reveal that for a constant perovskite thickness (a specific filtered spectrum), $V_{oc}$ and FF of the bottom sub-cell are decreased by increasing the thickness of the silicon absorber layer. On the other hand, for a constant thickness of the silicon absorber layer, increasing the perovskite thickness results in a slight reduction of $V_{oc}$ and FF of the bottom sub-cell. It is because the opportunity of charge carrier separation is decreased, by increasing the thickness of the absorber layer. On the contrary, the $J_{sc}$ of the bottom sub-cell is increased slightly when the silicon thickness is increased due to increasing light absorption and generation rate in the silicon layer. Then, $J_{sc}$ of the bottom sub-cell tends to saturate as the thickness of silicon approaches 120 micron. It is consisted with the results shown in Figure 6. On the other hand, at a constant silicon thickness, the Jsc of the bottom sub-cell is decreased by increasing the perovskite thickness, due to the higher light absorption in the perovskite top sub-cell. The behavior of PCE is not linear. At a constant perovskite thickness, PCE of the bottom sub-cell increases to a maximum value and then decreases slightly, by increasing the silicon thickness. While at a constant thickness of the silicon layer, PCE is decreased by increasing the perovskite thickness. In addition, maximum PCE values of the bottom sub-cell can be achieved for very low perovskite thicknesses varying in the range of 50 nm to 100 nm and the silicon thicknesses varying in the range of 20 μm to 80 μm. Moreover, the maximum values of $J_{sc}$ of the bottom sub-cell can be achieved when the perovskite thickness is lower than 100 nm and the silicon thickness is between 60 and 120 μm. In addition, maximum values of $V_{oc}$ and FF of the bottom sub-cell are achievable for very low thicknesses of silicon and perovskite absorber layers. Similar behaviors are observed for the case where BL method is used for calculating filtered spectra and the corresponding results are exhibited in figure S12.

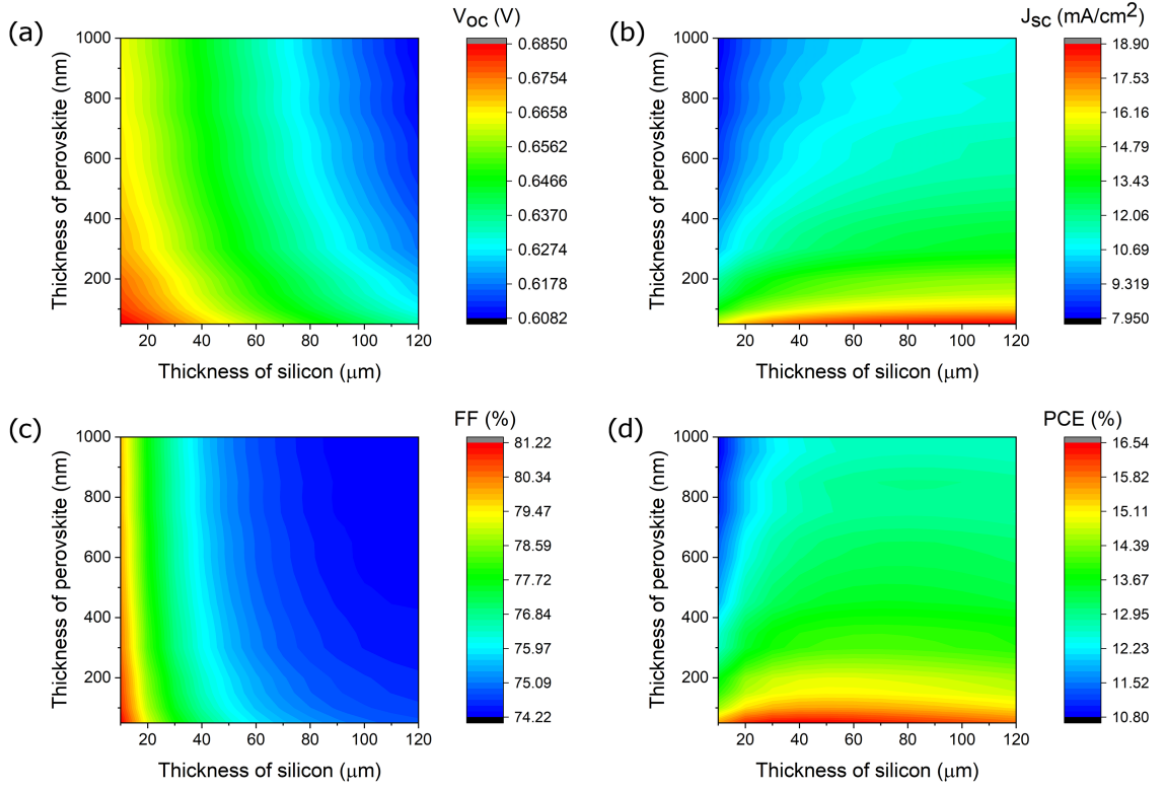

Figure 7: 2D counter plots of the performance parameters $V_{oc}$, $J_{sc}$, FF, and PCE of the bottom sub-cell, when illuminating by the filtered spectra calculated using TM method. The thickness of the silicon layer is altered from 10 to 120 μm in 12 steps and the thickness of the perovskite layer is changed from 50 to 1000 nm in 20 steps.

A comparison between the performance parameters $V_{oc}$, $J_{sc}$, FF and PCE of the silicon bottom sub-cell, when BL and TM methods are used for calculating the filtered spectra, are provided in Figure 8. The thickness of the silicon absorbing layer is 120 microns. Each data in Figure 8 is corresponded to a filtered spectrum that is calculated for a specific thickness of the perovskite layer. The observed behaviors are consisted with the results presented in Figure 7, except that the PV parameters of the bottom sub-cell in the case of using TM method are lower than the case where BL method is used. To describe precisely, the $J_{sc}$ values calculated by these two methods differ by almost 8 mA/cm$^2$ at the perovskite thickness of 50 nm and by 4 mA/cm$^2$ at the perovskite thickness of 1000 nm. Also, the difference between the values of PCE calculated by these two methods is varied between 2.3% at the perovskite thickness of 50 nm to 2.4% at the perovskite thickness of 1000 nm. Similarly, the difference between the values of $V_{oc}$ is approximately 0.023 V at the perovskite thickness of 50 nm and 0.023 V at the perovskite thickness of 1000 nm. Also, the values of FF calculated by these two methods differ by 0.45% at the perovskite thickness of 50 nm and by 0.40% at the perovskite thickness of 1000 nm. These differences that are more pronounced for PCE and $J_{sc}$ are ascribed to the additional optical losses that are incorporated in the formalism of TM method, as mentioned in sub-section 2-3.

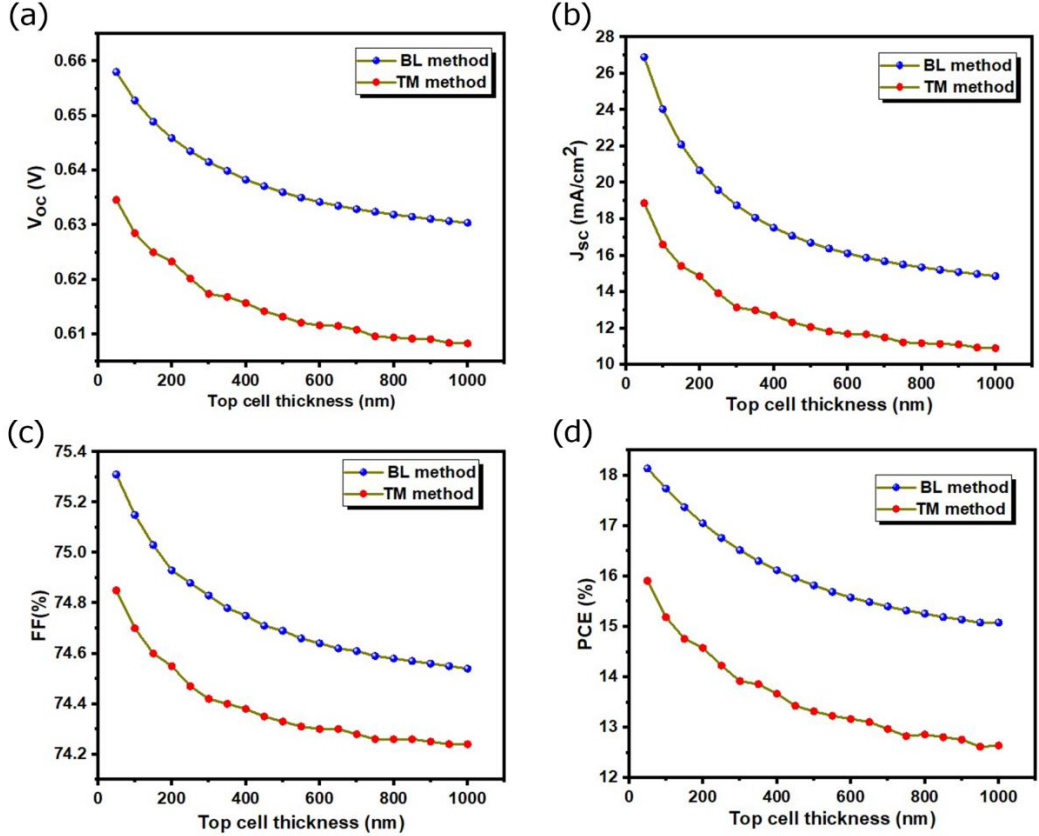

Figure 8: Comparing the performance parameters $V_{oc}$, $J_{sc}$, FF, and PCE of the bottom sub-cell using BL and TM methods. The silicon thickness is 120 μm, while the perovskite thickness is increased from 50 to 1000 nm.

Figure 9 presents the results of current matched analysis using TM and BL methods. The Jsc variation of the silicon bottom sub-cell as a function of the silicon thickness under the illumination of different filtered spectra along with the Jsc variation of the perovskite top sub-cell as a function of the perovskite thickness under the illumination of AM1.5G spectrum are depicted in Figure 9 (a) and (c) corresponded to the cases of using TM and BL methods, respectively. Using SCAPS-1D scripting features mentioned in the Supplementary Info of the reference [13], the optimized PV parameters of 2-T monolithic perovskite/Silicon TSC are calculated in both cases and the results are described in Table 1. The obtained optimized perovskite and silicon thicknesses for the case where TM method is used for calculating the filtered spectra are 158 nm and 80 μm, respectively. While the optimized perovskite and silicon thicknesses in the case where BL method is used are obtained as 246 nm and 40 microns, respectively. Figure 9 (b) and (d) shows the J–V curves of the standalone top sub-cell, standalone bottom sub-cell, bottom sub-cell under the illumination of the filtered spectrum and tandem cell at the current matched situation for the optimized thicknesses obtained in the cases of using TM and BL methods, respectively. It is observed that only one current is passed through top and bottom sub-cells, as was expected from the current matching technique. The current matching curves corresponding to the optimized thicknesses in the cases where TM and BL methods are used is depicted in Figure 9(e). Evidently, the $J_{sc}$ values of the silicon bottom sub-cell are lower when TM method is used. It is due to the increased optical losses that are incorporated in the formalism of TM method. Another noticeable point in Figure 9(e) is the difference between the optimized thicknesses of the perovskite absorbing layer predicted by these two methods (see the current matched points in Figure 9(e)).

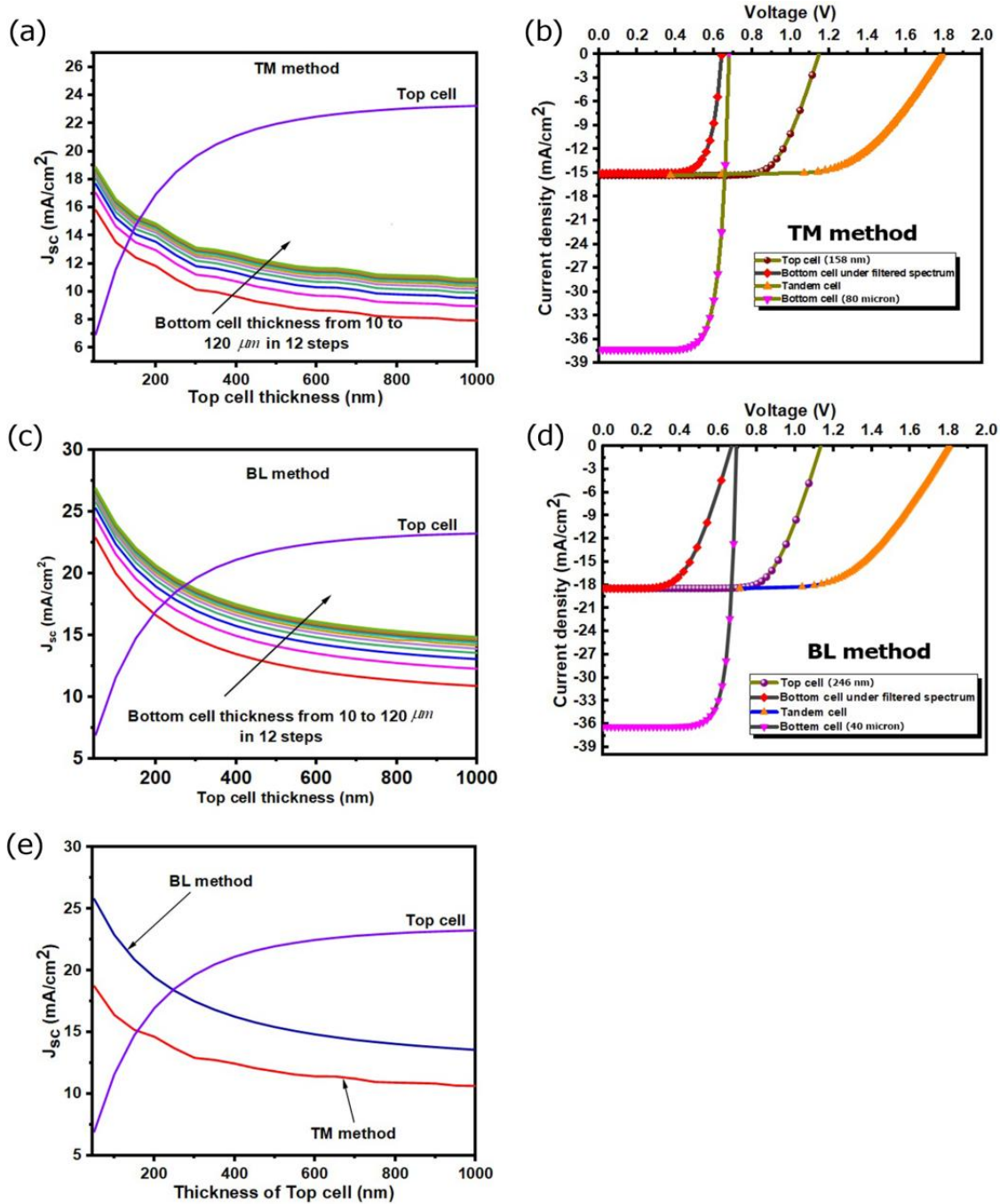

Figure 9: (a) Current matching curves for the case where TM method is used for calculating filtered spectra, (b) J–V curve of standalone top and bottom sub-cells, bottom sub-cell under the illumination of the filtered spectrum (calculated using TM method), and the tandem cell, (c) Current matching curves for the case where BL method is used for calculating filtered spectra, (d) J–V curve of standalone top and bottom sub-cells, bottom sub-cell under the illumination of the filtered spectrum (calculated using BL method), and the tandem cell, (e) Current matching curves corresponding to the maximum efficiency of tandem device, simulated using BL and TM methods.

Table 1. Performance parameters of the standalone top sub-cell, standalone bottom sub-cell, bottom sub-cell under filtered spectrum, and tandem solar cell using TM and BL methods.

| Optimized performance parameter | BL method | | | | TM method | | | |
|---|---|---|---|---|---|---|---|---|
| | Standalone top cell (246 nm) | Standalone bottom cell (40 µm) | Bottom cell under filtered spectrum | Tandem solar cell | Standalone top cell (158 nm) | Standalone bottom cell (80 µm) | Bottom cell under filtered spectrum | Tandem solar cell |
| PCE (%) | 14.45 | 19.86 | 17.06 | 21.08 | 12.58 | 19.43 | 14.72 | 17.82 |
| Jsc (mA/cm$^2$) | 18.42 | 36.36 | 18.45 | 18.42 | 15.17 | 37.33 | 15.09 | 15.17 |
| Voc (V) | 1.135 | 0.695 | 0.670 | 1.8 | 1.15 | 0.68 | 0.64 | 1.79 |
| FF (%) | 69.11 | 78.57 | 77.06 | 63.46 | 72.11 | 76.55 | 74.94 | 65.67 |

By comparing the detailed data presented in Table 1 for the optimized 2-T monolithic perovskite/Silicon TSC, it is clearly observed that the optimized thicknesses for the perovskite and silicon layers as well as the PV parameters depend on the method of calculating the filtered spectrum. Differences in the optimized thicknesses of the absorbing layers of top and bottom sub-cells are noteworthy in practical fabrication and any changes in the thicknesses of the fabricated layers can affect the performance parameters as well as the fabrication costs of tandem cells. In addition, increasing the number of layers will enhance the differences in the results obtained by using these two methods. This is due to the fact that the optical losses caused by light reflections and scatterings from grain boundaries and interfaces increase with the number of layers and their thicknesses. Also, the size of grain boundaries and scattering elements in the layers affect the light scattering that can be addressed using TM method. Hence, the BL method is unable to reveal the effects of the light scattering, interfacial reflections and reflections from the grain boundaries, except parasitic light absorption in the calculations. Given these details, using TM method will produce simulation findings that experimentalists can rely on for real-world applications.

### 3-2 Investigation of optical loss effects related to RJ

So far, the focus of our work has been on the effects of using TM and BL methods on the PV parameters of the bottom sub-cell as well as the tandem configuration. Herein, we use TM and BL methods to investigate the optical loss effects related to the RJ in the SCAPS-1D device simulator. Two perovskite/Si TSCs that differ in their RJs are studied: the first one with Spiro-OMeTAD (HTL)/n$^{++}$-Si hybrid RJ and the second one with Spiro-OMeTAD (HTL)/ITO/n$^{++}$-Si RJ. First, it is needed to optimize two different bottom sub-cells under the AM1.5G light irradiation. In the first configuration that the RJ is formed by HTL/n$^{++}$-Si stack, we optimize the thickness of n$^{++}$-Si layer to boost the efficiency of the silicon bottom sub-cell. In the second configuration that the RJ is formed by HTL/ITO/n$^{++}$-Si stack, we optimize the ITO thickness. The thicknesses of n-Si and p$^{++}$-Si in both configurations are 80 micron and 20 nm, respectively. Figure 10(a) represents the PCE variation of the silicon bottom sub-cell in the first configuration against the thickness of n$^{++}$-Si layer. The obtained result show that the optimum n$^{++}$-Si thickness of 50 nm can achieve the highest efficiency of 19.54% along with the $V_{oc}$ of 0.68 V, $J_{sc}$ of 37.53 mA/cm$^2$, and FF of 76.55%, under the AM1.5G light irradiation. Then, the silicon bottom sub-cell in the second configuration is studied and the PCE dependence of the silicon bottom sub-cell on the variation of ITO thickness is investigated. The thickness of $n^{++}$-Si layer in this configuration is assumed as 100 nm. According to Figure 10(b), the ITO thickness of 10 nm is the optimum value that attains the highest efficiency of 19.82% along with the $V_{oc}$ of 0.6810 V, $J_{sc}$ of 38.41 mA/cm$^2$, and FF of 75.74% for the silicon bottom sub-cell.

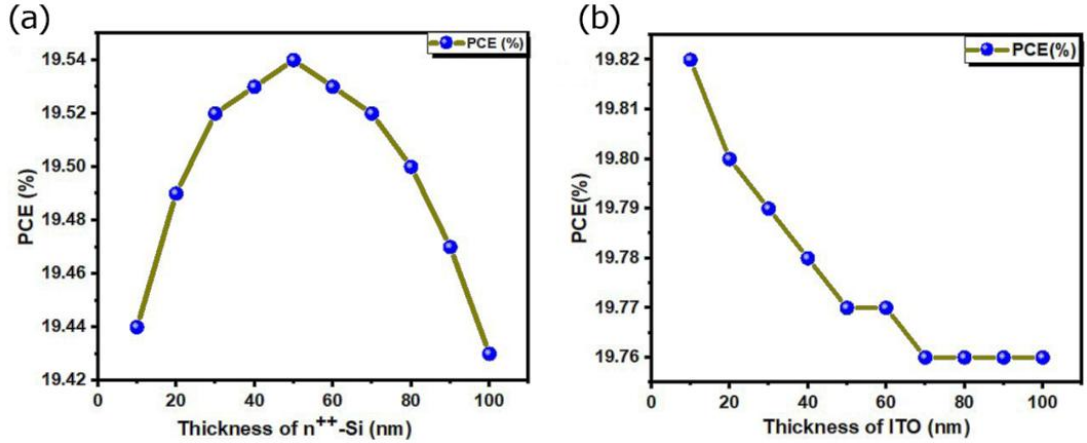

Figure 10: The PCE dependence of the silicon bottom sub-cell as a function of: (a) thickness of $n^{++}$-Si layer in the configuration with HTL/ $n^{++}$-Si, (b) thickness of ITO layer in the configuration with HTL/ITO/ $n^{++}$-Si.

Figure 11 shows the effects of introducing ITO recombination layer to the Spiro-OMeTAD/$n^{++}$-Si stack on the PV parameters of the silicon sub-cell. Both TM and BL methods are used for the calculation of different filtered spectra corresponding to different perovskite thicknesses. The thickness of silicon absorber layer is 120 μm. As it was expected, an improvement in the performance parameters of the silicon bottom sub-cell by the addition of ITO recombination junction is observed only when the TM method is integrated with SCAPS-1D.

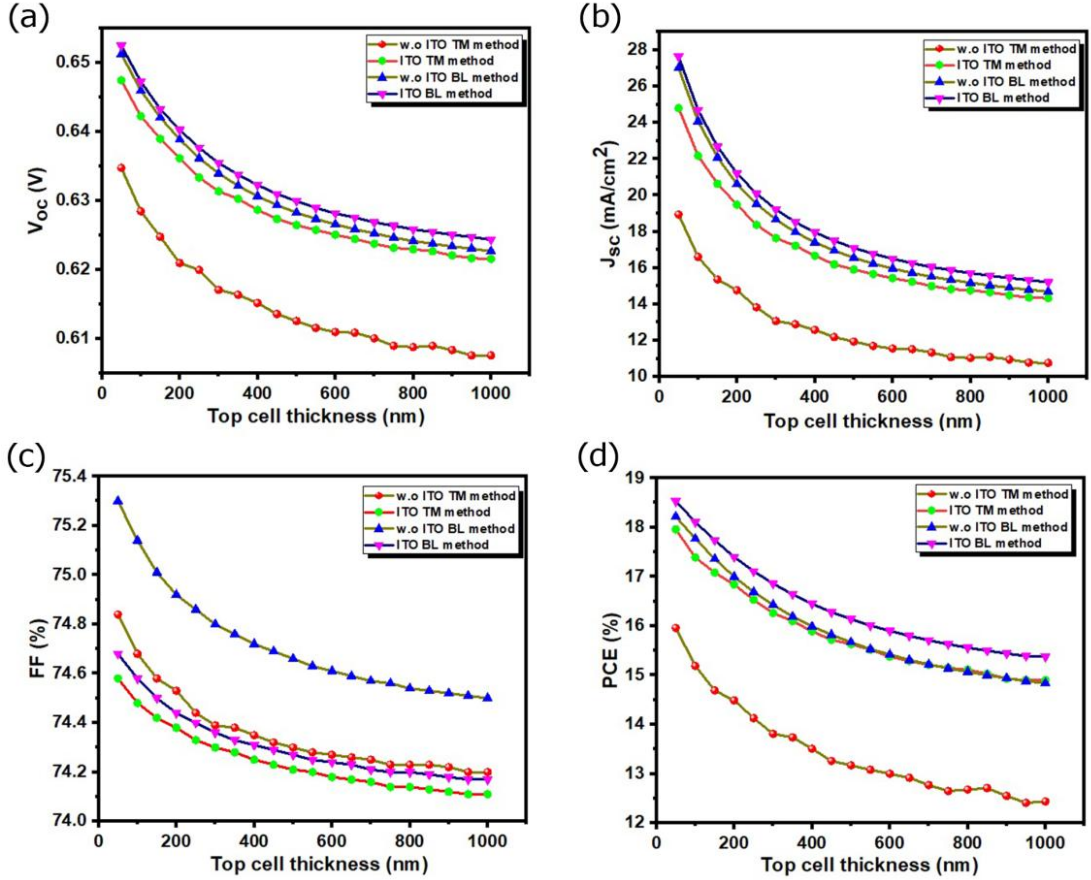

Figure 11: The dependence of the PV parameters: (a) $V_{oc}$, (b) $J_{sc}$, (c) FF and (d) PCE of the silicon bottom sub-cell on the modification of RJ, by sandwiching ITO between Spiro-OMeTAD and $n^{++}$-Si. The filtered spectra are calculated by using both TM and BL methods.

Figure 12 shows the results of current match analysis for two different perovskite/Silicon TSCs with different RJs. Two top figures depict the results for the case where TM method is used for calculating the filtered spectra, while two bottom figures show the results corresponded to the case where BL method is used. As shown in Figure 12(a), using TM method, the current matched point between the perovskite top sub-cell and the silicon bottom sub-cell is enhanced from 15.31 mA/cm$^2$ to 18.41 mA/cm$^2$, when the ITO layer is added to the Spiro-OMeTAD/$n^{++}$-Si hybrid RJ. On the other hand, the optimized perovskite thickness is increased from 161 nm to 245 nm, when the silicon thickness is constant and equals 120 micron. Figure 12(b) illustrates the J–V curves of the standalone top sub-cell, bottom sub-cell under the illumination of filtered spectrum and the tandem cell at two different current matched situations that are corresponded to two different RJs. The filtered spectra are calculated using TM method. As it was expected, the matched current passes through the top and bottom sub-cells are increased by introducing ITO recombination layer to the Spiro-OMeTAD/$n^{++}$-Si RJ. However, using BL method, the optimized perovskite thickness and the matched current shows no difference, as it is evident from Figure 12(c) and (d). Table 2 elucidates the detailed values of the performance parameters $V_{oc}$, $J_{sc}$, FF, and PCE of the standalone top sub-cell (STC), the bottom sub-cell under the illumination of the filtered spectrum (FBC) and the tandem solar cell (TSC), when the current matched situation is established for two different configurations of the bottom sub-cell. The filtered spectra are calculated using TM method. The thickness of the silicon absorber layer is 120 micron. The obtained results show that introducing ITO as the recombination layer to the Spiro-OMeTAD/$n^{++}$-Si hybrid RJ improves the performance of the tandem solar cell device. Our simulation findings are consisted with the discussions presented in [25]. Hence, integrating TM method with the SCAPS-1D device simulator makes possible the investigation of RJs in 2-T monolithic TSCs.

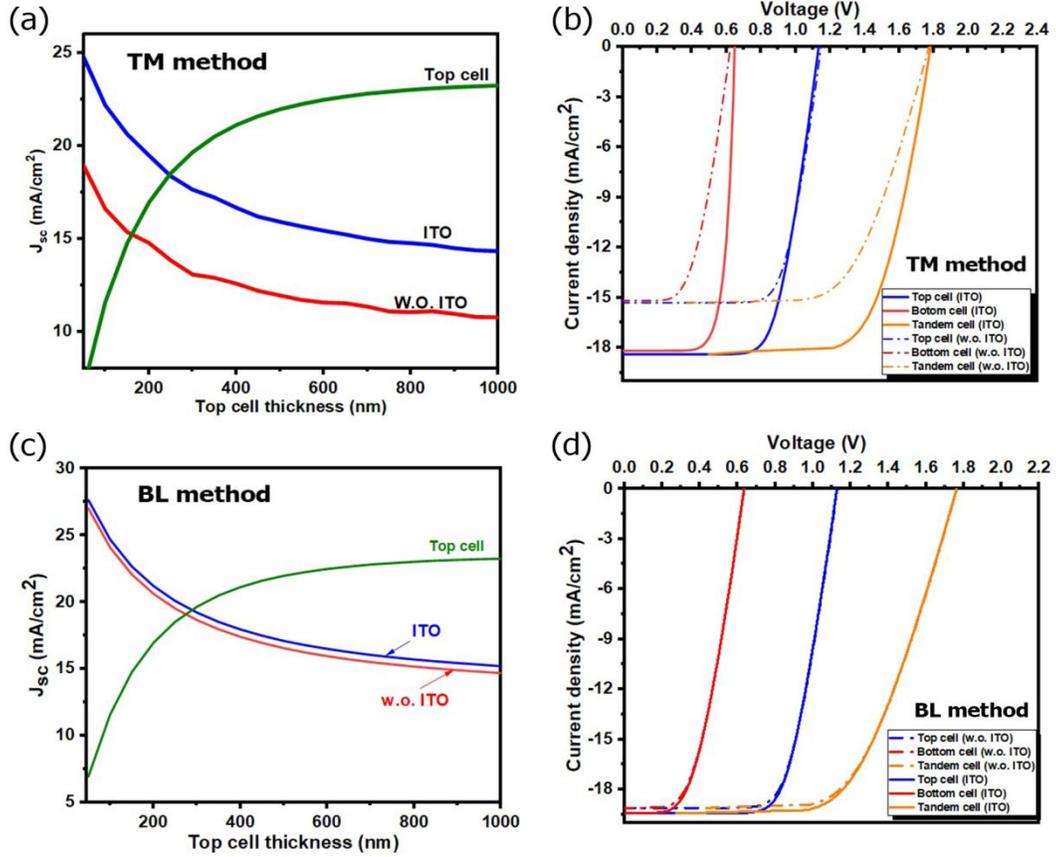

Figure 12: Current matching curves corresponding to the maximum efficiency of the tandem device as well as J–V curves of standalone top sub-cell, bottom sub-cell under the illumination of filtered spectrum and tandem cell with two different RJs formed by Spiro-OMeTAD/$n^{++}$-Si (w.o. ITO) and Spiro-OMeTAD/ITO/$n^{++}$-Si (ITO): (a, b) TM method, and (c, d) BL method are used for calculating the filtered spectra.

Table 2. Performance parameters of the standalone top cell (STC), bottom cell under filtered spectrum (FBC), and tandem solar cell (TSC) using the TM method for calculating the filtered spectra. The thickness of silicon absorber layer is 120 micron. Thickness values mentioned in the parenthesis are related to the perovskite absorber layer. Data are corresponded to two different RJs formed by Spiro-OMeTAD/$n^{++}$-Si (w.o. ITO) and Spiro-OMeTAD/ITO/$n^{++}$-Si (ITO):

| Optimized performance parameters | Standalone top cell | | Bottom cell under filtered spectrum | | Tandem solar cell | |
|---|---|---|---|---|---|---|
| | ITO (245 nm) | Without ITO (161 nm) | ITO (245 nm) | Without ITO (161 nm) | ITO | Without ITO |
| PCE (%) | 14.44 | 12.67 | 16.33 | 14.49 | 23.10 | 19.81 |
| Jsc (mA/cm$^2$) | 18.40 | 15.31 | 18.46 | 15.28 | 18.40 | 15.31 |
| Voc (V) | 1.13 | 1.15 | 0.630 | 0.624 | 1.77 | 1.81 |
| FF (%) | 69.13 | 71.98 | 74.34 | 74.59 | 71.06 | 71.48 |

## 3-1 Per-layer optical loss analysis using TM method

In this sub-section, using TM method, we analyze the performance of PSC from an optical point of view. Figure 13(a) exhibits the contribution of each layer to the total absorbance, when stacked on each other in the PSC. The respective spectra are obtained by calculating the total electric filed inside each layer, as described in [31]. The hybrid RJ is assumed as Spiro-OMeTAD/$n^{++}$-Si. The thicknesses of FTO, $SnO_2$, CsFAMA and Spiro-OMeTAD layers are assumed constant and equal to 100 nm, in order to focus on the material dependent behavior of light through transmitting the perovskite solar cell. It is observed that FTO and perovskite layers have more contribution to the total absorbance, in comparison with the other layers. Figure 13(b) provides a schematic illustration of the amount of light absorbed in each layer and transmitted from the PSC. The specified wavelengths 410 nm, 480 nm, 530 nm, 570 nm, 610 nm, 630 nm and 800 nm are chosen to represent the violet, blue, green, yellow, orange, red and NIR spectra. The percentages in the layers and bellow the Spiro-OMeTAD denote the light absorbance and transmittance at the selected wavelengths, respectively. From Figure 13(b), it is possible to estimate the optical loss and the fraction of light transmitted from the PSC at the specified wavelengths. In addition, Figure 13(c) depicts the total transmittance, total reflectance and total absorbance of PSC that are calculated from equations 10, 11 and 12, respectively. Obviously, the amount of light lost through reflection is significant for the light with wavelengths above 600 nm. This optical loss, when the PSC is implemented as the top sub-cell in a perovskite/Si tandem configuration, will reduce the performance of bottom sub-cell. Therefore, modifying the interface between the perovskite top sub-cell and the silicon bottom sub-cell in a 2-T monolithic perovskite/Silicon TSC is a very important issue.

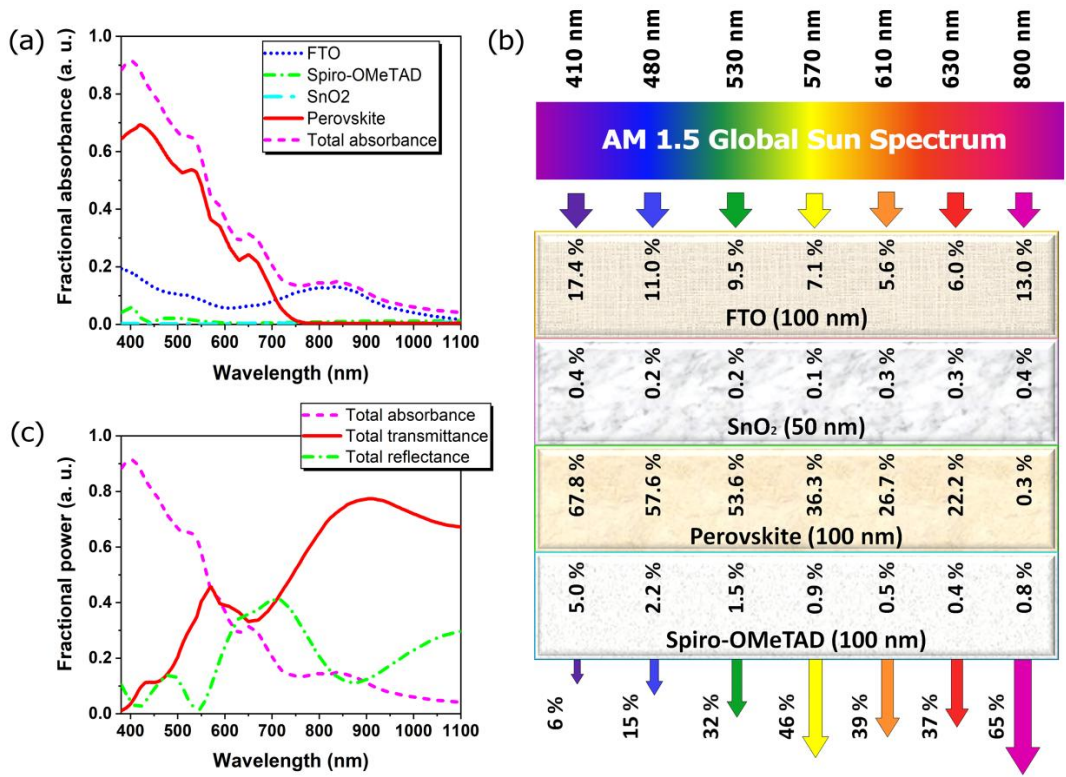

Figure 13: (a) Total absorbance and the fraction of light absorbed in each layer, when they are stacked on each other to form the perovskite solar cell, (b) Schematic illustration of the perovskite top sub-cell under the AM1.5G light irradiation. The percentage of light absorbed in the layers along with the light transmitted from the cell is denoted in the figure. The specified colors represent the wavelengths of 410 nm, 480 nm, 530 nm, 570 nm, 610 nm, 630 nm and 800 nm for the incident light. (c) Total absorbance, transmittance and reflectance spectra of light, when it passes across the perovskite top sub-cell.

Figure 14 (a), (b) and (c) compares the total absorbance, reflectance and transmittance spectra of the perovskite top sub-cell, when the ITO recombination layer is introduced to the RJ. It is deduced from Figure 14(a) that the absorbance of PSC shows no significant change, by the addition of ITO between Spiro-OMeTAD and $n^{++}$-Si. However, Figure 14(b) reveals that the reflection loss at NIR wavelengths is greatly reduced, when the RJ consists of ITO recombination layer. According to equation 5, the light reflectance at each layer depends on the refractive indices of its adjacent layers. As depicted in Figure 5, introducing ITO decreases the refractive index mismatch when light exits the PSC and results in the reduction of reflection losses. Therefore, more light is transmitted from the PSC at NIR wavelengths, as clearly observed in Figure 14(c). Obtained results suggest that integrating TM method with SCAPS-1D makes the optimization and evaluation of RJs in 2-T monolithic TSCs possible. Based on these spectra presented in Figure 14, the layers can be selected in such a way that the reflection and parasitic absorption losses in the layers become minimized. The present study can pave the way for further theoretical investigations on optimizing RJs, via SCAPS-1D.

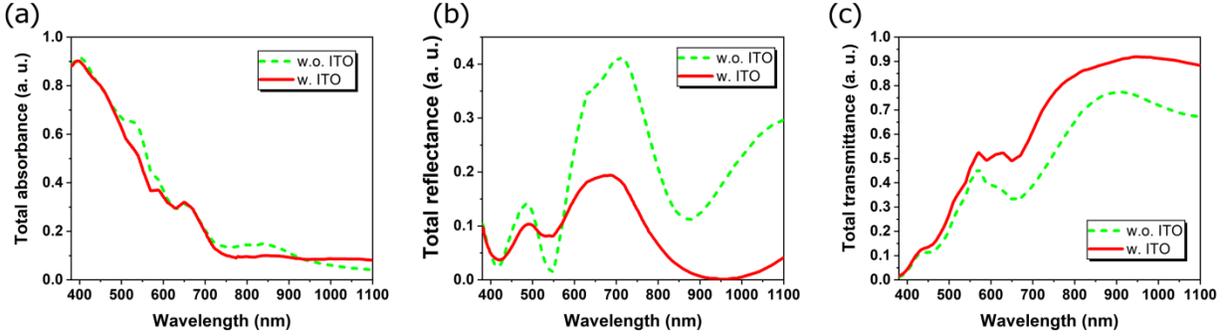

Figure 14: (a) Total absorbance, (b) Total reflectance and (c) Total transmittance spectra of the perovskite top sub-cell. The thicknesses of FTO, SnO$_2$, CsFAMA and Spiro-OMeTAD are 100, 50, 100 and 100 nm, respectively.

## 4- Conclusion

In this paper, the TM method was integrated into SCAPS-1D to investigate a 2-terminal monolithic perovskite/Silicon TSC from both electrical and optical point of views. The corresponding results were compared with the case in which BL method was integrated into SCAPS-1D. It was demonstrated that using BL method for calculating the filtered spectra transmitted from the perovskite top sub-cell results in an overestimation of the photovoltaic properties and the corresponding SCAPS-1D simulations are unable to fully address all phenomena that light encounters in each layer and interfaces. Therefore, using BL method can only give a poor prediction of optimized thicknesses, layer configurations in TSCs and their corresponding photovoltaic metrics. Our theoretical investigation revealed a reduction of ~4% in the efficiency of tandem device, when the optical losses are addressed correctly in the SCAPS-1D by using TM method. In addition, the effects of ITO thin add-layer sandwiched between Spiro-OMeTAD and n$^{++}$-Si in the interconnection layer, on cell efficiency were studied using TM and BL methods. The results confirmed that BL method is unable to include optical effects at the recombination junction and returns a negligible change on cell efficiency for ITO-included device. However, an improvement of tandem device efficiency from 19.81% to 23.10% was observed with TM method for the same configuration. The results demonstrated that using TM method integrated into SCAPS-1D can pave the way for further accurate optoelectronic investigations on optimizing RJs in TSCs.

**Acknowledgment**

The authors acknowledge Vice-Presidency of Sci. & Tech., Iran and Center for National Macro Technology Projects (grant number 11.69206) for financial support. MK thanks Prof. H. Nadgaran, Physics Dept., Shiraz Uni. for constructive discussions and his group support to complete this project.

**Conflicts of interest**

There are no conflicts of interest to declare.

# Supplementary Information

# Integrating transfer matrix method into SCAPS-1D for addressing optical losses and per-layer optical properties in perovskite/Silicon tandem solar cells

**Peymaneh Rafieipour\*, Aminreza Mohandes, Mohammad Moaddeli, Mansour Kanani**

[1]Department of Materials Science and Engineering, School of Engineering, Shiraz University, Shiraz, Iran

*Corresponding author: prafieepoor@yahoo.com*

**Content:**

### S1: Optical study of the perovskite top sub-cell

The fraction of light that is absorbed in, reflected and transmitted from the $SnO_2$, CsFAMA perovskite and Spiro-OMeTAD layers are depicted in Figures S1 (a), S1 (b) and S1 (c), respectively. The light intensity spectra are calculated by using the TM method. In the calculations, the thicknesses of 50 nm, 550 nm and 300 nm are used for $SnO_2$, CsFAMA and Spiro-OMeTAD layers, respectively. As it is observed in Figure S1 (a), $SnO_2$ has an excellent transparency and is a suitable ETL to be deposited on top of the CsFAMA absorber layer. The absorbance, transmittance and reflectance spectra of the CsFAMA perovskite are shown in Figure S1 (b). It is then deduced that the 550 nm thick CsFAMA absorber layer has good transparency for wavelengths above 750 nm. Figure S1(c) shows the transparency and the fraction of light lost in the Spiro-OMeTAD HTL. It is inferred from this figure that the parasitic absorption in the Spiro-OMeTAD is significant for the light with near-IR wavelengths. One way to reduce this optical loss is reducing the thickness of Spiro-OMeTAD. Therefore, the fraction of light absorbed in the Spiro-OMeTAD as a function of its thickness is plotted in Figure S1 (d). The results demonstrate that the parasitic light absorption in the Spiro-OMeTAD at near-IR wavelengths is reduced, when its thickness is decreased from 500 nm to 100 nm.

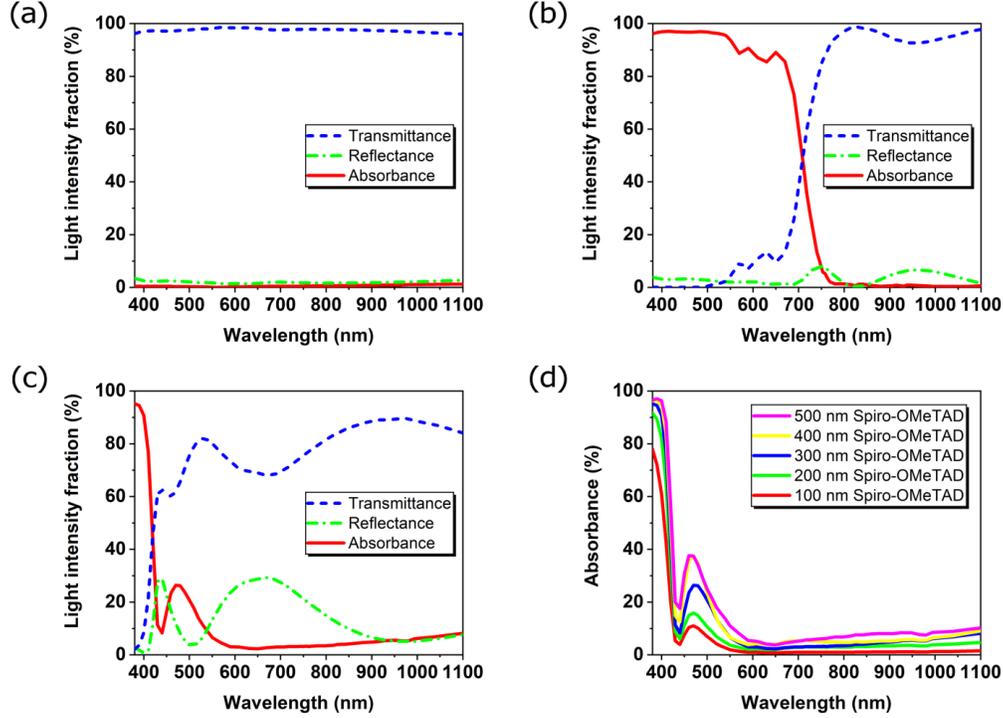

Figure S1: Transmittance, absorbance and reflectance spectra of: (a) SnO$_2$, (b) Perovskite and (c) Spiro-OMeTAD layers. Their thicknesses are 50 nm, 550 nm and 300 nm, respectively. (d) The light absorbance in the Spiro-OMeTAD layer as a function of its thickness.

### S2: SCAPS relations

Table S1: Fundamental equations used during SCAPS 1-D simulation [1]

| | | |
|---|---|---|
| Model of optical in SCAPS | $N_{ph}(\lambda, x) = N_{ph0}(\lambda).T_{front}.\exp(-\alpha(\lambda).x) \times$ $\left(\frac{1+R_{back}(\lambda).\exp(-2(d-x).\alpha(\lambda))}{1-R_{back}(\lambda).R_{int}.\exp(-2d.\alpha(\lambda))}\right)$ $G(\lambda, x) = \alpha(\lambda).N_{ph}(\lambda, x)$ $G(x) = \int_{\lambda_{min}}^{\lambda_{max}} G(\lambda, x)d\lambda$ | Where $N_{ph}(\lambda, x)$ is the photon flux at any position in the layer, $N_{ph0}(\lambda)$ incident photon flux, $T_{front}$ is the front contact transmission, $\alpha(\lambda)$ is the absorption coefficient, $R_{int}$ is the interior reflection at the front contact, $R_{back}(\lambda)$ is the back contact reflection, d is the layer thickness, x is the layer position. $G(\lambda, x)$ is the generation rate of electron-hole pair, $\lambda$ is the wavelength. |
| 1-D semiconductor equation | $\frac{d^2\rho}{dx^2} = \frac{-q(p - n + N_D^+ - N_A^- + p_t - n_t)}{\varepsilon_0 \varepsilon_r}$ $-\frac{dJ_n}{dx} - R_n + G = \frac{dn}{dt}$ $-\frac{dJ_p}{dx} - R_p + G = \frac{dp}{dt}$ | Where q is the electron charge, $\rho$ is the electrostatic potential, p, n, n$_t$ are the free hole, free electron, trapped electron, and trapped hole, respectively. $N_D^+$ and $N_A^-$ are ionized donor and acceptor doping concentration. $\varepsilon_0$ is vacuum permittivity and $\varepsilon_r$ is the permittivity of material, $J_n$ is the |
| | $$J_n = -\frac{\mu_n n}{q}\frac{dF_n}{dx}$$ $$J_p = +\frac{\mu_p p}{q}\frac{dF_p}{dx}$$ | electron current density, $J_p$ is the hole current density. $R_n$ is the total recombination rate for electron, $R_p$ is the total recombination rate for hole. $\mu_n$ and $\mu_p$ are mobility of electron and hole, respectively. $F_n$ is quasi fermi level of electron a quasi fermi level of hole. In the steady-state condition, $\frac{dn}{dt} = \frac{dp}{dt} = 0$. |
| Defect of Gaussian | $[E_t - \frac{w_G}{2}E_C, E_t + \frac{w_G}{2}E_C]$ $$N_t(E) = N_{peak}.\exp\left[-\left(\frac{E-E_t}{E_c}\right)^2\right]$$ $$N_{tot}(N_{peak}) = E_C.N_{peak}$$ | Where, $E_t$ is the trap energy level, $E_C$ is the characteristic energy, $w_G$ width of Gaussian energy distribution, $N_t(E)$ is the defect density, $N_p$ the energy density at the peak of energy distribution, $N_{tot}(N_{peak})$ is the total defect density integrated over all energies. |

### S3: Standalone top and bottom sub-cells

Band gap energy ($E_g$ (eV)), electron affinity ($\chi$ (eV)), dielectric permittivity ($\varepsilon_r$), conduction band effective density of state ($N_C$ (cm$^{-3}$)), valence band effective density of state ($N_V$ (cm$^{-3}$)), electron mobility ($\mu_n$ ($cm^2/V.s$)), hole mobility ($\mu_p$ ($cm^2/V.s$)), shallow donor concentration ($N_D$ (cm$^{-3}$)), shallow acceptor concentration ($N_A$ (cm$^{-3}$)), and defect density in the bulk of material ($N_t$ (cm$^{-3}$)) are the parameters used in the simulation and are listed in Table . All parameters of SnO$_2$ and Spiro-OMeTAD are exactly the same as article [2], and all parameters for perovskite absorber layer are taken from article [2], except for $N_A$ which obtained by fitting with the experimental current density-voltage (J-V) curve. The values of $N_{ts}$, the defect density at the interface, at the rear interface (CsFAMA layer/HTL) is $4.5 \times 10^8\ cm^{-2}$ to achieve the recombination velocities of 45.0 cm/s. The work function for front (FTO) and rear (Au) contacts is 4.5 eV and 5.1 eV, respectively. The absorption coefficient (α) is computed by the formula α=$A_\alpha$(h$\nu$-$E_g$)$^{1/2}$, in which $A_\alpha$ (1.0×10$^5$ cm$^{-1}$ eV$^{-1/2}$) is the Pre-factor and is calculated by fitting with experimental J-V curve, h$\nu$ is the photon energy, and $E_g$ is the bandgap energy. $R_s$ is denoted as series resistance with the unit of ($\Omega.cm^2$).

Table S2: Basic parameters needed for PSC simulation. All values are extracted from ref (Mohandes et al., 2021) otherwise the reference is mentioned.

| Parameter | SnO$_2$ | Perovskite | Spiro-OMeTAD |
|---|---|---|---|
| Thickness (nm) | 50 | 700 | 300 |
| $N_A$ (cm$^{-3}$) | 0 | $1 \times 10^{15}$[fitting] | $1 \times 10^{18}$ |
| $N_D$ (cm$^{-3}$) | $1 \times 10^{19}$ | 0 | 0 |
| $E_g$(eV) | 4.04 | 1.57-1.60 | 3.0 |
| $\chi$ (eV) | 4.09 | 3.9 | 2.45 |
| $\epsilon_r$ | 9 | 32 | 3.0 |
| $\mu_n$ ($cm^2/V.s$) | 240 | 50 | $2 \times 10^{-4}$ |
| $\mu_p$ ($cm^2/V.s$) | 25 | 50 | $2 \times 10^{-4}$ |
| $N_t$ (cm$^{-3}$) | $1 \times 10^{18}$ | $1.28 \times 10^{16}$ (Bu et al., 2017) $1.47 \times 10^{16}$ (Bu et al., 2017) | $1 \times 10^{15}$ |
| $N_C$ (cm$^{-3}$) | $2.2 \times 10^{18}$ | $2.2 \times 10^{18}$ | $2.2 \times 10^{18}$ |
| $N_V$ (cm$^{-3}$) | $1.8 \times 10^{19}$ | $1.8 \times 10^{19}$ | $1.8 \times 10^{19}$ |

The optical filter of CsFAMA% is tabulated in Table , which reveals the optical reflection at various wavelengths at the interface of glass/FTO. The defect's energy level is located in the center of the Eg and is dispersed in a Gaussian distribution with a characteristic energy of 0.1 eV. The defect type is neutral and the electron and hole capture cross sections are $2 \times 10^{-14}$ cm$^2$ (Minemoto et al., 2019; Minemoto & Murata, 2015), except for the CsFAMA layer. The defect density, $N_t$, is related to the capture cross section of the hole and electron as in the following formula:

$$N_t = \frac{1}{\sigma_{n,p} \times \tau_{n,p} \times v_{th}} \quad \text{(eq. S1)}$$

where $v_{th}$ is the thermal velocity of electrons and holes, $\tau_{n,p}$ is the carrier life time of electrons and holes and is equal to 210 ns. $\sigma_{n,p}$ is the capture cross section of electrons and holes. The values of $\tau_{n,p}$ are available in sup. Info. of the ref(Bu et al., 2017). Therefore, the capture cross section of the hole and electron for the CsFAMA layer are $3.245 \times 10^{-17}$ cm$^2$ and $3.727 \times 10^{-17} cm^2$, respectively.

Table S3: Optical filter of CsFAMA% against wavelength, optical reflection

| Wavelength (nm) | Filter (%) | Wavelength (nm) | Filter (%) | Wavelength (nm) | Filter (%) |
|---|---|---|---|---|---|
| 0 | 8 | 530 | 7 | 740 | 7 |
| 340 | 34 | 540 | 7 | 750 | 6 |
| 350 | 35 | 550 | 7 | 755 | 6 |
| 365 | 32 | 560 | 6 | 770 | 6 |
| 370 | 30 | 580 | 6 | 780 | 60 |
| 375 | 23 | 585 | 6 | 785 | 65 |
| 380 | 20 | 590 | 5 | 790 | 43 |
| 390 | 20 | 610 | 5 | 795 | 43 |
| 410 | 13 | 620 | 5 | 800 | 43 |
| 450 | 9 | 630 | 5 | 10000 | 42 |
| 460 | 8 | 640 | 6 | | |
| 480 | 8 | 660 | 8 | | |
| 490 | 7 | 680 | 8 | | |
| 500 | 7 | 700 | 7 | | |
| 505 | 7 | 720 | 7 | | |
| 510 | 7 | 730 | 7 | | |

As presented in Table , the perovskite absorber layer, called CsFAMA layer, is p-type and the hole and electron defect densities in the CsFAMA layer are observed to be $1.47 \times 10^{16}$ and $1.28 \times 10^{16} \, cm^{-3}$, respectively [3]. It causes the electron and hole carrier lifetimes to be 210 ns and the electron and hole carrier diffusion lengths, $L_n$ and $L_p$ to be 5.2 and 5.2 µm, respectively.

First, we reproduced the experimental J-V characteristic of the PSC with 15.52% efficiency mentioned in the literature [3]. The simulation was employed with the parameters considered in Table and Table , as well as Table which represents the fitted parameters with experimental J-V curve for the rear contact interface.

Table S4: Rear interface defect density which is calculated by fitting.

| Parameters | CsFAMA layer/HTL |
|---|---|
| Type of defect | Neutral |
| $\sigma_{n,p}$ (cm$^2$) | $1.0 \times 10^{-14}$ |
| The distribution of Energy | Single |
| Reference for defect energy level $E_t$ | Above the highest $E_v$ |
| Energy with respect to Reference (eV) | 0.35 |
| $N_t$ (cm$^{-2}$) | $4.5 \times 10^8$ |

The device is optimized by the perovskite absorber layer thickness variation from 50 to 1000 nm in 20 steps, while the thickness of HTL is 300 nm. Figure (a) depicts the calibration of J-V curve. Whereas, the calibration of EQE curve is shown in Figure (b), under the AM1.5G illumination.

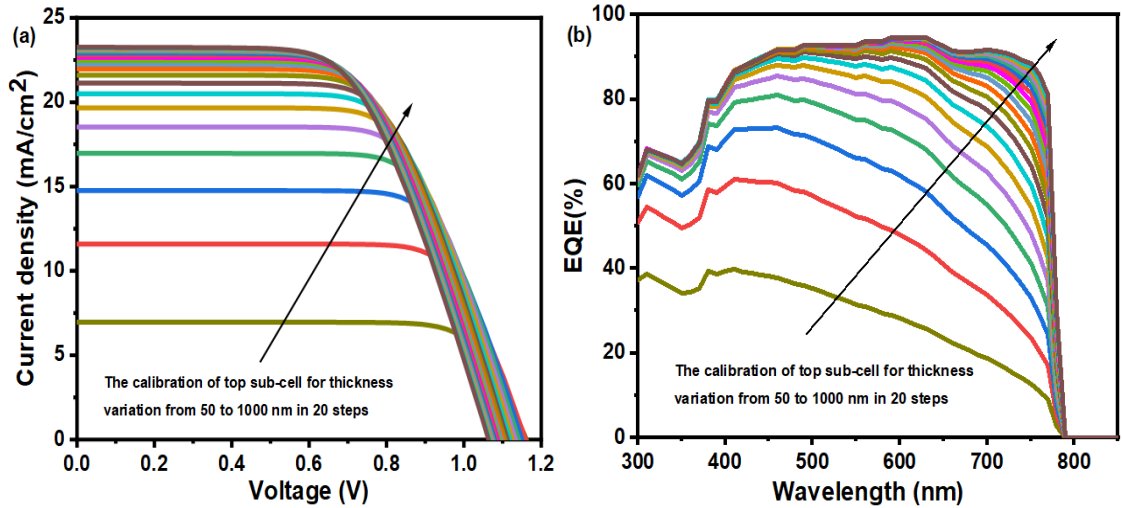

Figure S2: Calibration of the top sub-cell with the perovskite thickness variation from 50 to 1000 nm in 20 steps: (a) J-V curve, (b) EQE curve under the illumination of AM1.5G spectrum. The thickness of HTL is 300 nm.

As shown in Figure S2(a), we find the optimized thickness of 550 nm for the perovskite absorber layer and 300 nm for the HTL, under the AM1.5 G spectrum. Figure (a) depicts the J-V curve, which represents the calibration of the top sub-cell with a 550 nm perovskite absorber layer and 300 nm HTL. Whereas, the EQE curve calibrated with the same thicknesses and under the illumination of AM1.5G spectrum is shown in Figure (b). The results show that the performance parameters of PCE = 15.68%, $V_{oc}$ =1.09 V, $J_{sc}$ = 22.25 mA/cm$^2$, and FF = 64.21% with $R_s$ = 9.97 Ω.cm$^2$ are gained.

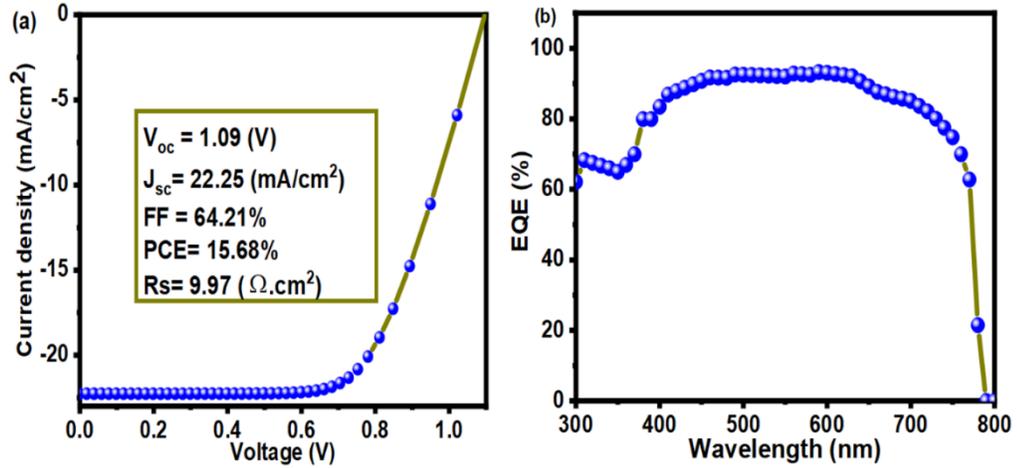

Figure S3: The calibration of the top sub-cell with an optimized thickness of 550 nm for the CsFAMA layer and 300 nm for HTL: (a) J-V curve, (b) EQE curve under the illumination of AM1.5G spectrum.

Figure (a) shows the result of J-V curve when the thickness of the perovskite absorber layer and the Spiro-OMeTAD are 500 and 100 nm, respectively. Figure (b) displays the EQE curve for the aforementioned thicknesses and under the AM1.5G illumination. A PCE of 15.68%, $V_{oc}$ of 1.10 V, $J_{sc}$ of 21.97 mA/cm$^2$, and FF of 64.79%, with $R_s$ of 9.97 $\Omega$.cm$^2$ are obtained from the J-V curve.

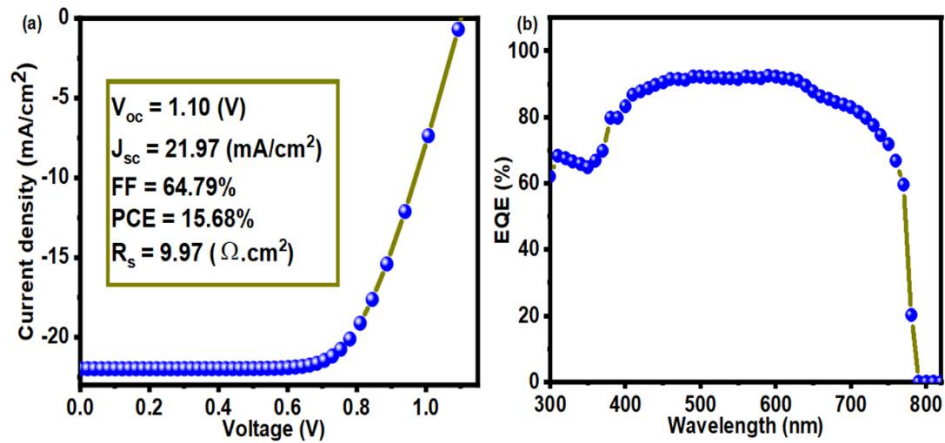

Figure S4: The calibration of the standalone top sub-cell: (a) J-V curve, and (b) EQE curve under the AM1.5G light irradiation. The thickness of CsFAMA layer is 500 nm and thickness of HTL is 100 nm.

The energy band diagrams for the perovskite top sub-cell are attained by device simulation and illustrated in Figure (a) and (b). To comprehend the production and direction of the electron-hole pair motion, data are found in both dark and light conditions. Furthermore, the presence of interface produces an appropriate electric field to separate and collect the charge carriers and avoids the movement of the same counterpart, as presented in Figure (b).

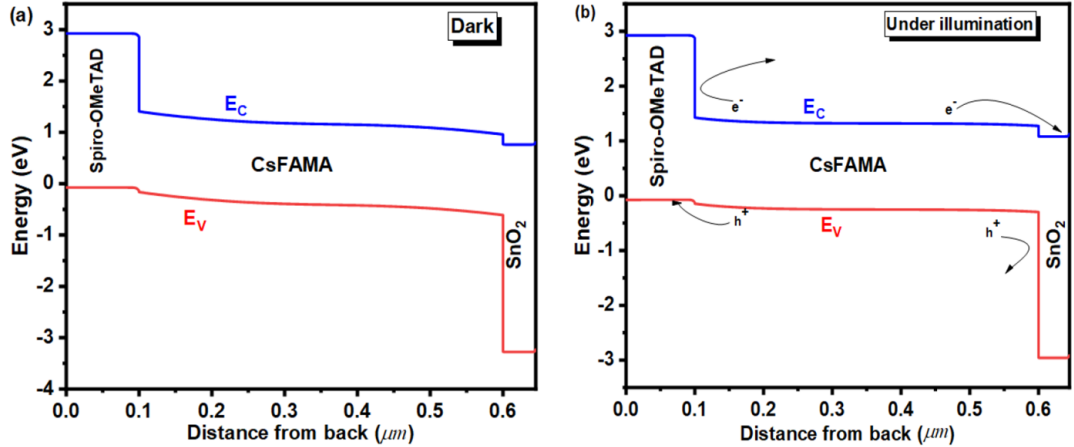

Figure S5: Energy band diagram of perovskite solar cell under: (a) dark condition, and (b) AM1.5G illumination.

The performance parameters as a function of the thickness variation can be comprehended in Figure S. For an increase in thickness up to 700 nm, $J_{sc}$ increases to 22.81 mA/cm$^2$ and after that almost remains constant. This increase is because of the enhanced absorption and the generation rate. Nevertheless, the inverse changes of $V_{oc}$ with respect to thickness have occurred owing to the decreased electric field strength across the thick perovskite layer. The excitons created by the absorption of photons are unable to go over the barrier potential which is the depletion layer. This may give rise to greater recombination rate of charge carriers and leads to a decrease in the $V_{oc}$. Consequently, the reduction of the electric field reduces the possibility of charge carrier separation and diminishes the $V_{oc}$ and FF of the PSC. Increasing the series resistance of thick perovskite layers is another factor that helps to reduce FF. The influence of $V_{oc}$, $J_{sc}$, and FF on PCE is found, which confirms that an increase in $J_{sc}$ dominates the decrease in $V_{oc}$ and FF. In addition, PCE increases to 15.53% with increasing the thickness up to 400 nm and after that almost remains unchanged. The various experimental studies have already displayed that the performance of solar cells is highly dependent on the morphology of the perovskite absorber layer, which has a direct effect on the photocurrent carrier lifetime and diffusion length [6, 7].

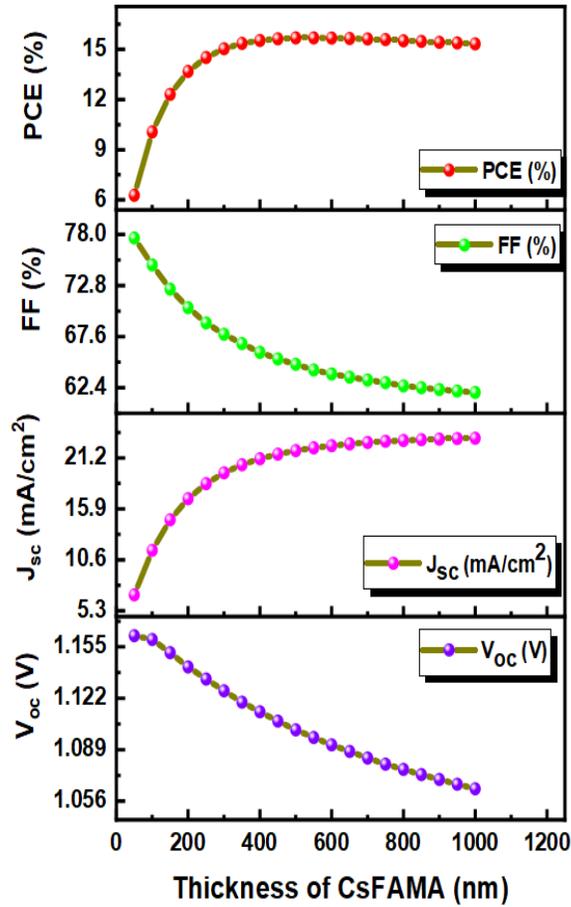

Figure S6: The performance parameters $V_{oc}$, $J_{sc}$, FF, and PCE of the perovskite top sub-cell as a function of the perovskite (CsFAMA) thickness variation, under the AM1.5 Glight irradiation.

The silicon solar cell, which is comprised of $p^{++}$-Si/n-Si/$n^{++}$-Si and has been calibrated before (Amri et al., 2021) is used in our study. To identify the charge carrier dynamics for the silicon bottom sub-cell, energy band diagrams are drawn in Figure (a) and (b). In addition, the presence of the interface creates a suitable electric field for the separation and collection of charge carriers and the suppression of similar counterpart motion, as shown in Figure (b).

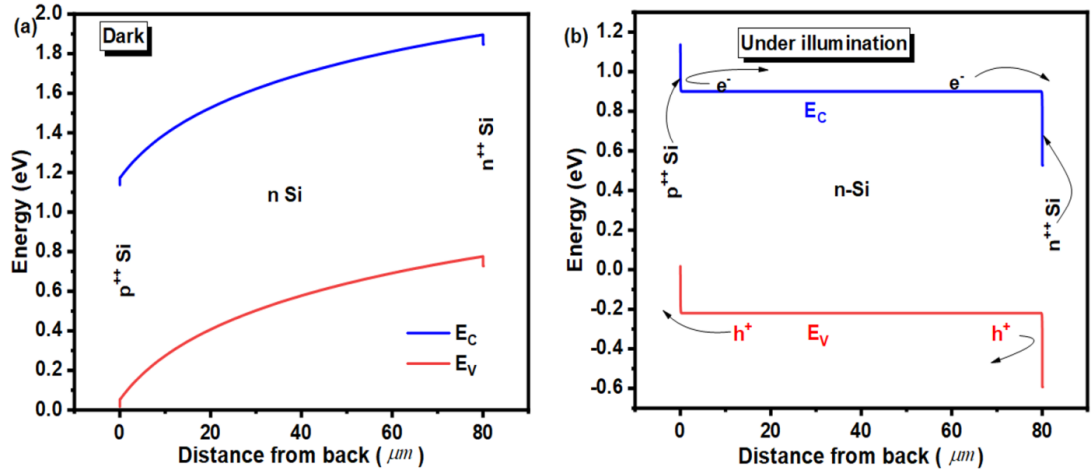

Figure S7: Energy band diagram of Si bottom sub-cell under: (a) dark condition and (b) AM1.5G illumination.

Now, we change the thickness of the silicon absorber layer from 10 to 120 μm and study the J–V and EQE curves of the bottom sub-cell. The obtained results for different thicknesses of the n-Si layer are presented in Figure S8.

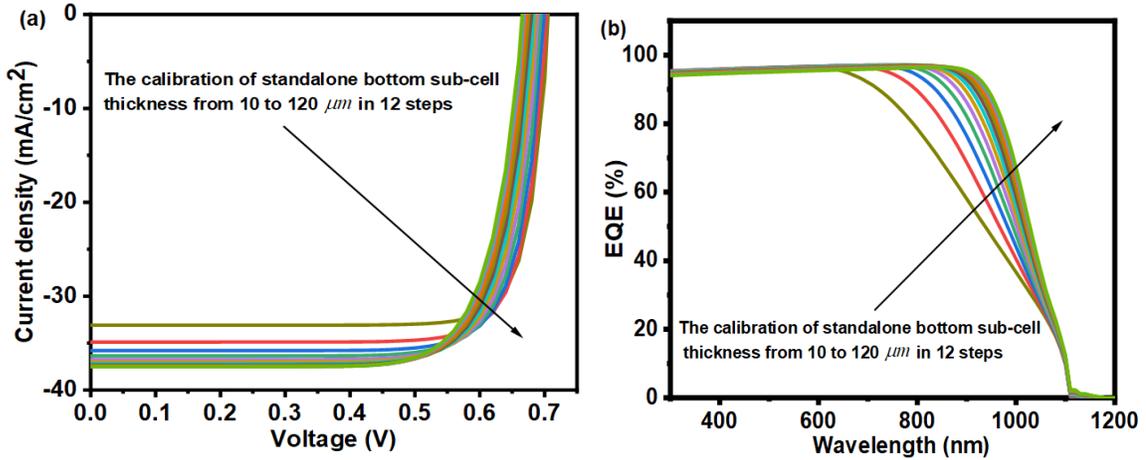

Figure S8: (a) J-V curves, and (b) EQE curves of the bottom sub-cell as a function of the silicon thickness variation from 10 to 120 μm, in 12 steps. The standalone calibration is done under the AM1.5G illumination.

Table S5: Elementary parameters for silicon simulation (Burgelman et al., 2000). (Amri et al., 2021)

| Parameters | $n^{++}$-si | n-si | $p^{++}$-si |
|---|---|---|---|
| Thickness ($\mu m$) | 0.1 | 80 | 0.02 |
| $N_A$ (cm$^{-3}$) | 0 | 0 | $5.0 \times 10^{19}$ |
| $N_D$ (cm$^{-3}$) | $10^{22}$ | $10^{14}$ | 0 |
| $E_g$ (eV) | 1.12 | 1.12 | 1.12 |
| $\chi$ (eV) | 4.05 | 4.05 | 4.05 |
| $\epsilon_r$ | 11.9 | 11.9 | 11.9 |
| $\mu_n$ ($cm^2/V.s$) | $1.04 \times 10^3$ | $1.04 \times 10^3$ | $1.04 \times 10^3$ |
| $\mu_p$ ($cm^2/V.s$) | $4.2 \times 10^2$ | $4.2 \times 10^2$ | $4.2 \times 10^2$ |
| $N_C$ (cm$^{-3}$) | $2.8 \times 10^{19}$ | $2.8 \times 10^{19}$ | $2.8 \times 10^{19}$ |
| $N_V$ (cm$^{-3}$) | $2.6 \times 10^{19}$ | $2.6 \times 10^{19}$ | $2.6 \times 10^{19}$ |

The performance parameters of the Si bottom sub-cell as a function of the Si thickness variation can be comprehended in Figure S9.

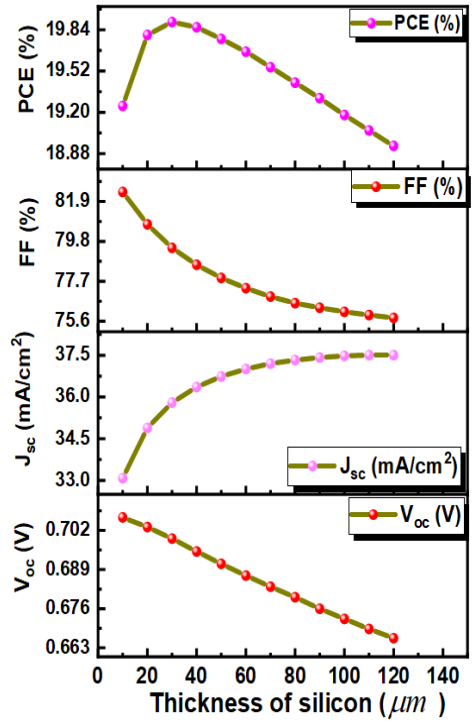

Figure S9: The performance parameters $V_{oc}$, $J_{sc}$, FF, and PCE of the bottom sub-cell as a function of the thickness of the silicon absorber layer.

## S4: Calculated filtered spectra using BL and TM methods

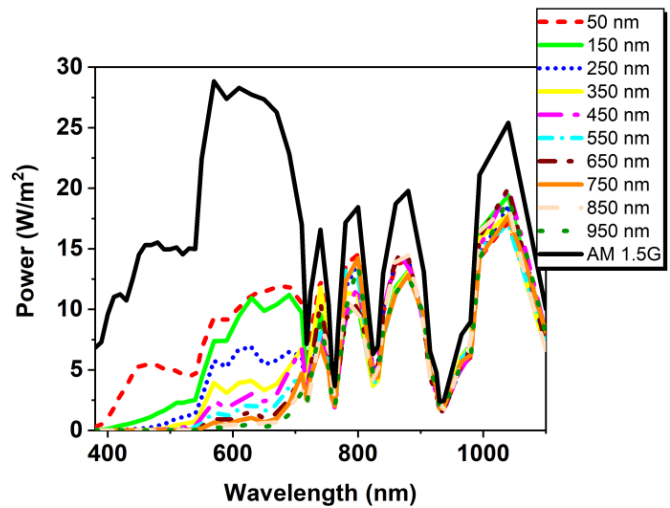

Figure 15: AM1.5G and filtered spectra. Different filtered spectra are corresponded to different perovskite thicknesses and are calculated by using the transfer matrix (TM) method.

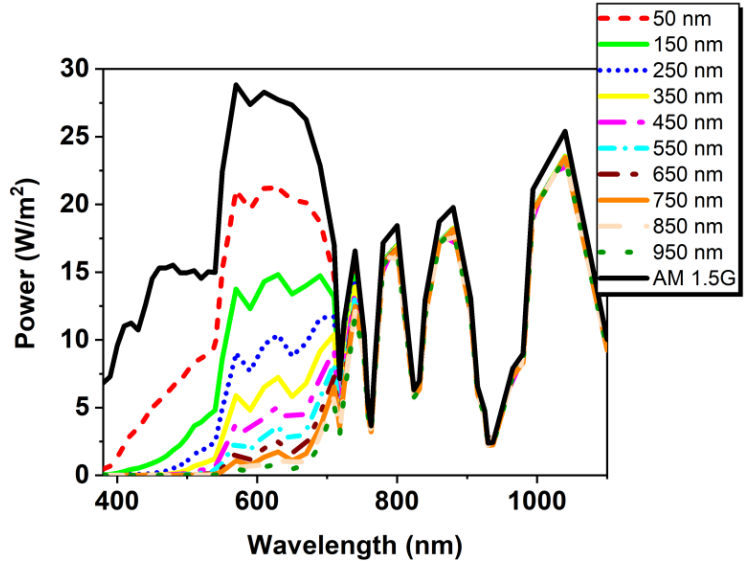

Figure 16: AM1.5G and filtered spectra. Different filtered spectra are corresponded to different perovskite thicknesses and are calculated by using the Beer-Lambert (BL) method.

## S5: Detailed values of $J_{sc}$ of the bottom sub-cell

Table S6: $J_{sc}$ values of the bottom sub-cell corresponding to different silicon absorber layer thicknesses. The silicon bottom sub-cell is studied under different filtered spectra (using the TM method) corresponding to different perovskite thicknesses. $J_{sc}$ values are reported for the minimum (10 µm), maximum (500 µm), and optimized thicknesses of the silicon absorber layer.

| Perovskite thickness (nm) | Silicon thickness (µm) | $J_{sc}$ of bottom sub-cell (mA/cm$^2$) |
|---|---|---|
|  | 10 | 15.82 |
| 50 | **120** | **18.88 (max)** |
|  | 500 | 14.75 |
|  | 10 | 12.43 |
| 150 | **130** | **15.44 (max)** |
|  | 500 | 12.18 |
|  | 10 | 10.15 |
| 300 | **140** | **13.18 (max)** |
|  | 500 | 10.32 |
|  | 10 | 8.49 |
| 700 | **140** | **11.54 (max)** |
|  | 500 | 9.14 |
|  | 10 | 8.11 |
| 900 | **150** | **11.16 (max)** |
|  | 500 | 8.85 |
|  | 10 | 7.92 |
| 1000 | **140** | **10.96 (max)** |
|  | 500 | 8.65 |

## S6: Counter plots of the photovoltaic parameters of the bottom sub-cell

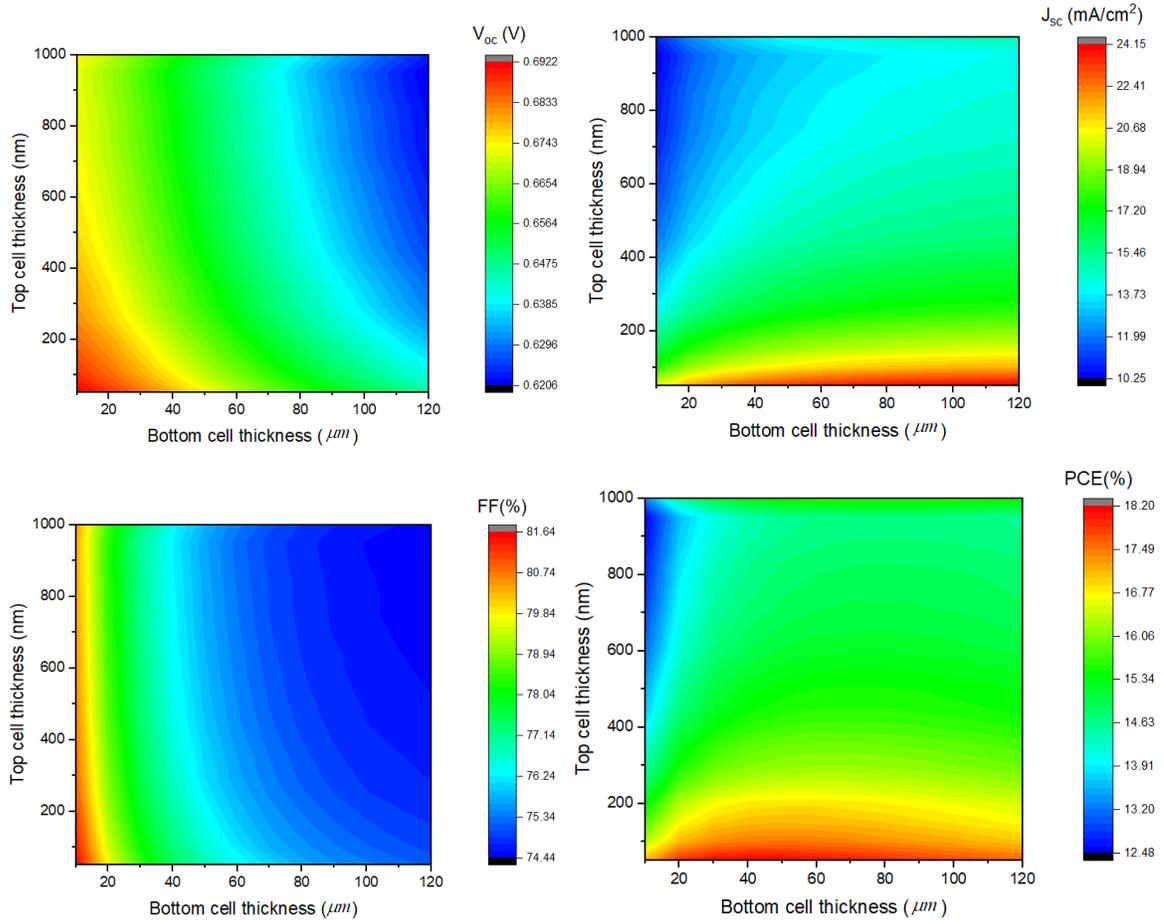

Figure 17: The 2D counter plots of the performance parameters $V_{oc}$, $J_{sc}$, FF, and PCE of the bottom sub-cell. The filtered spectra are calculated by using the Beer-Lambert method. The thickness of silicon layer is varied from 10 to 120 in 12 steps and the thickness of the perovskite layer is changed from 50 to 1000 nm in 20 steps.

## S7: References